\documentclass[ACS,STIX1COL]{WileyNJD-v2}

\articletype{Full Article}%

\received{ }
\revised{ }
\accepted{ }

\usepackage{graphicx}
\usepackage{amsmath,amsfonts,amssymb}
\usepackage{tabu}
\usepackage{pdflscape}
\usepackage{makecell}
\usepackage{booktabs}
\usepackage{pdfpages}
\usepackage{epstopdf}
\usepackage{multirow}
\usepackage{enumitem}
\usepackage{wrapfig}
\usepackage{textcomp}
\usepackage{tablefootnote}
\usepackage{blindtext}
\usepackage{subfig}
\DeclareGraphicsExtensions{.jpg,.png,.eps}

\begin{document}

\title{An ultrasonography-based approach for optical diagnostics and phototherapy treatment strategies}

\author[1]{Akhil Kallepalli*}
\author[2]{James Halls}
\author[3]{David B. James}
\author[3]{Mark A. Richardson}
\authormark{Kallepalli \textsc{et al}}

\address[1]{\orgdiv{School of Physics and Astronomy}, \orgname{University of Glasgow}, \orgaddress{\state{Glasgow}, \country{United Kingdom}}}

\address[2]{\orgdiv{Department of Radiology}, \orgname{The Great Western Hospital}, \orgaddress{\state{Swindon}, \country{United Kingdom}}}

\address[3]{\orgdiv{Centre for Electronic Warfare, Information and Cyber}, \orgname{Cranfield University}, \orgaddress{\state{Defence Academy of the UK, Shrivenham}, \country{United Kingdom}}}

\corres{* Direct all correspondence to AK \\ \email{Akhil.Kallepalli@glasgow.ac.uk}}

\presentaddress{* Optics Group, School of Physics and Astronomy, Kelvin Building, University of Glagsow, Glasgow G12 8QQ}

\abstract{Currently, diagnostic medicine uses a multitude of tools ranging from ionising radiation to histology analysis. With advances in piezoelectric crystal technology, high-frequency ultrasound imaging has developed to achieve comparatively high resolution without the drawbacks of ionising radiation. This research proposes a low-cost, non-invasive and real-time protocol for informing photo-therapy procedures using ultrasound imaging. We combine currently available ultrasound procedures with Monte Carlo methods for assessing light transport and photo-energy deposition in the tissue. The measurements from high-resolution ultrasound scans is used as input for optical simulations. Consequently, this provides a pipeline that will inform medical practitioners for better therapy strategy planning. While validating known inferences of light transport through biological tissue, our results highlight the range of information such as temporal monitoring and energy deposition at varying depths. This process also retains the flexibility of testing various wavelengths for individual-specific geometries and anatomy.}


\jnlcitation{\cname{%
\author{A. Kallepalli}, 
\author{J. Halls}, 
\author{D. B. James}, and 
\author{M. A. Richardson}} (\cyear{2021}), 
\ctitle{An ultrasonography-based approach for optical diagnostics and phototherapy treatment strategies}, \cjournal{J. of Biophotonics}, \cvol{2021;xx:1--x}.}

\maketitle

\footnotetext{\textbf{Abbreviations:} HFUS, High frequency ultrasound; MC, Monte Carlo; PDT, Photodynamic therapy}

\section{Introduction} \label{sec:introduction}
Amongst the many radiology tools for diagnostics, ultrasound imaging has evolved into an indispensable modality due to its real-time results and non-ionising nature. It is and continues to be a key addition to diagnostic medicine \textit{in vivo}. In this research, we provide a strategy for individual-specific diagnostics to plan more effective photodynamic therapy protocols. To this end, individual analysis of tissue geometries is necessary. We use high-frequency ultrasound imaging (HFUS) for assessing the depth and thickness of the layers of the skin for building optical models. We utilise proprietary and open-source Monte Carlo (MC) platforms to achieve the optical modelling of light propagation in these models. As a result, we propose a relatively inexpensive and rapid assessment technique to plan phototherapy and photo-modulation strategies and assess the impact of using light therapy. This method can be added as a step before treatment to improve the effectiveness of optical therapies for treating conditions such as skin cancer, for instance. We envisage that this research will translate to using ultrasound imaging to construct 3D models of the anatomy, assign suitable optical properties and simulate the treatment strategies in a point-of-care approach. 

Models have previously been used to understand the laser-induced injury to skin \cite{Jean2013}. Such studies provide an understanding of light-induced damage yet use approximated/assumed dimensions for the tissue layers. Additionally, optical properties may not be individual- or site-specific. Most optical therapy applications require the quantification of optical properties \cite{Bashkatov2016}. It has also been found that the post-PDT recurrence rate of basal cell carcinoma is higher in comparison to surgical excision \cite{Cohen2016}. This is clearly due to remnant cancerous cells in the skin that proliferate after the procedure. Our study provides a solution to all the above challenges with accurate anatomical dimensions, individual-specific optical properties and the ability to accurately judge the impact of the therapy before treatment. 

\section{Background} \label{sec:background}
Years of medical research and innovation has resulted in the progression of ultrasound imaging to achieve better resolution and enhanced point-of-care capabilities \cite{Rallan2003}. In the last two decades, significant advances in piezoelectric crystals, probes and computational capabilities have enhanced ultrasound imaging with higher frequency systems and high-resolution imaging \cite{Rallan2003,Kleinerman2012}. This has allowed improved assessment of superficial body tissues, such as the skin. Generally, probes operating at frequencies greater than 15 MHz for dermatological studies \cite{Barcaui2015,Ketterling2017,Wortsman2013} (>13.5 MHz in some cases \cite{Kleinerman2012}) are considered to perform high-frequency ultrasound imaging (HFUS). Modern HFUS systems can provide sufficient resolution for a detailed assessment with recognition of individual skin layers and subcutaneous structures. The depth penetration is inversely proportional to the ultrasound probe frequency. A 7.5 MHz probe can assess subcutaneous tissues and lymph nodes to a depth of approximately 4 cm whilst probes above 50 MHz would only typically allow assessment of the epidermis, without the penetration to assess deeper skin layers. 

The second key element of this research is modelling light transport in biological tissue. Clinical experiments can be challenging because of costs, trials and ethics regulations. The applicability of methods and medications to people is the final frontier for a medical breakthrough. When using optical technologies, this is not very different. May this be remission-inducing laser exposures for destroying cancerous cells or dermatological cancer treatments, initial testing is not directly done on humans. This has ushered in the gold standard of light transport modelling, Monte Carlo (MC) simulations. Guided by probability and the knowledge of the optical properties of tissues, possible or likely outcomes are generated. This, done repeatedly, provides a realistic understanding of the problem. 

The MC technique is a robust method for simulating the interactions between light and tissue. The interaction of photons at each instance in the tissue is statistically sampled to track the variations to the optical flux and accounts for absorption, reflection, refraction and scattering events. Each photon is tracked until either its termination (due to absorption) or it exits the boundary limits of the model. After performing this simulation many times over, the overall distribution of all the paths yields a dependable approximation of reality. Before the popularisation of MC methods through Monte Carlo for multi-layered tissues (MCML) \cite{Wang1995}, models of absorption and distribution of energy in tissues were presented by Wilson and Adam (1983) \cite{Wilson1983}. However, MCML found general acceptance and applicability in simulating photon propagation. The scattering and absorption in biological tissues were widely researched \cite{Flock1989,Kumari2011,Prahl1989,Patwardhan2005}. MC methods were also used to determine the optical properties of tissues and chromophores \cite{Nishidate2004,Simpson1998,Shimada2001,Palmer2006,Meglinskii2001,Bjorgan2015}, investigating light interaction with blood vessels, simulation of pulse oximetry and heart rate detection \cite{Nilsson1998,AzorinPeris2007,Gan2011}, cancer detection \cite{Palmer2006a,Zhu2012} and modelling current and novel methods such as photoacoustic imaging \cite{Paltauf2018}. MC methods have been further developed with the implementation of GPU-based computations and modified for fast perturbation in turbid media \cite{Funamizu2014,Sassaroli2004}. These are only a few of the many examples in biomedical optics where MC simulations are today a standard for comparison with experimental work. 

\section{Methods} \label{sec:methods}
This article combines ultrasound imaging data, individual-specific optical properties and MC simulation-based assessment of the models to present a methodology for informing therapeutic strategies. Naturally, they form the main parts of the methodology. The MC simulations require geometry of the tissues and optical properties. The geometry is acquired using ultrasound imaging while the optical properties are decided using models and properties obtained from an extensive review of published literature. It is also important to state at this stage that optical properties that are specific to individuals can be included in this modality when combined with established spectroscopic techniques \cite{Sandell2011}. 

\subsection{Ultrasound Imaging}
All participants were scanned with a Samsung RS80A ultrasound system at the Great Western Hospital (Swindon, UK) by a single radiologist with over 10 years of experience. The ultrasound system includes a high-frequency 18 MHz linear transducer (probe), a visualisation and processing system, combined with network storage. Using the collected ultrasound data, we reliably measure the thickness of the tissue layers for each participant. 

\begin{figure}[h]
\centering
\includegraphics[width=0.75\columnwidth]{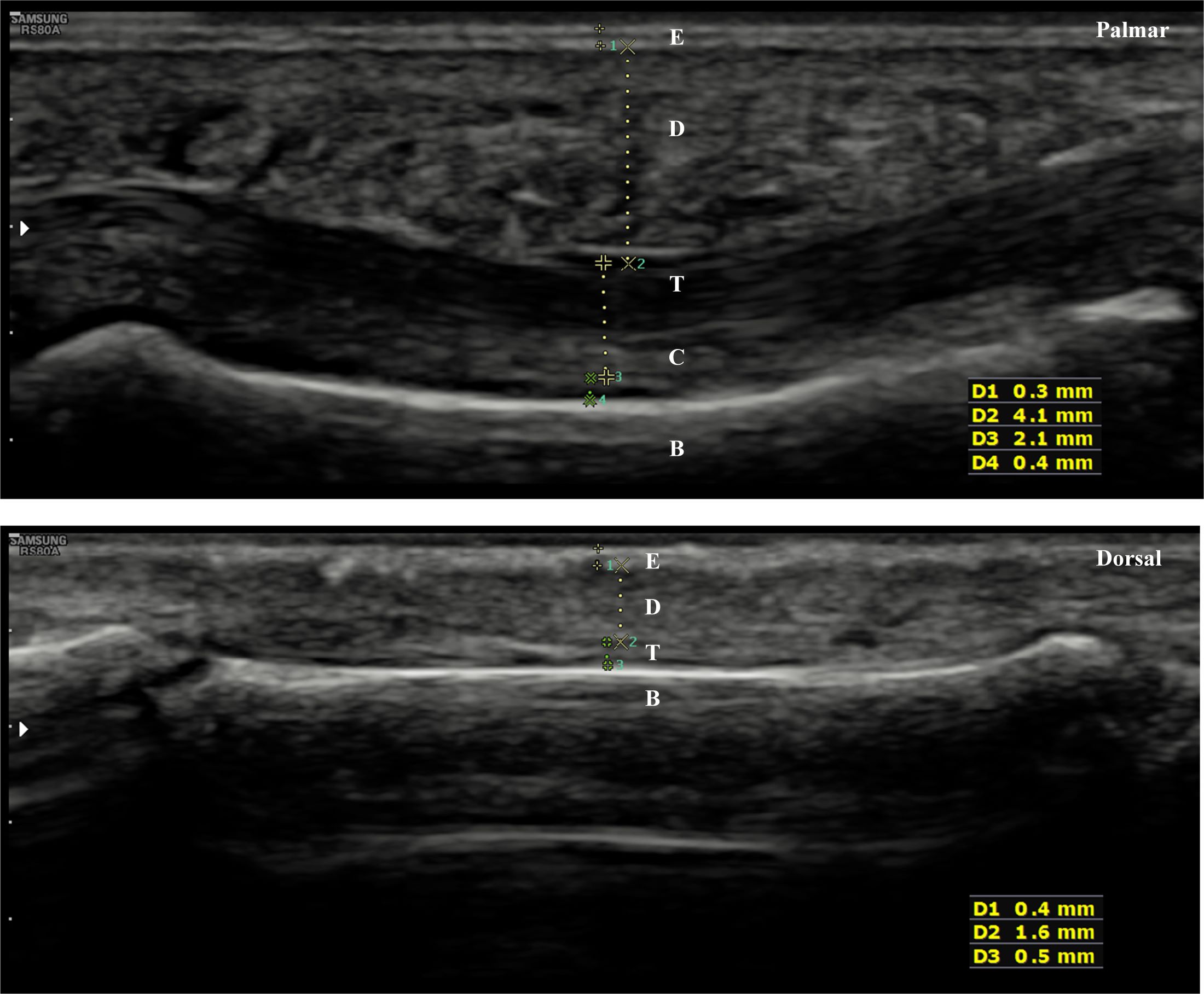}
\caption{Longitudinal orientation of the probe allowed imaging along the length of the finger, quantifying the anatomical entities. Both the palmar and dorsal sides of the finger are measured to complete the finger structure.}
\label{fig:US_Long}
\end{figure}

\begin{figure}[h]
\centering
\includegraphics[width=0.75\columnwidth]{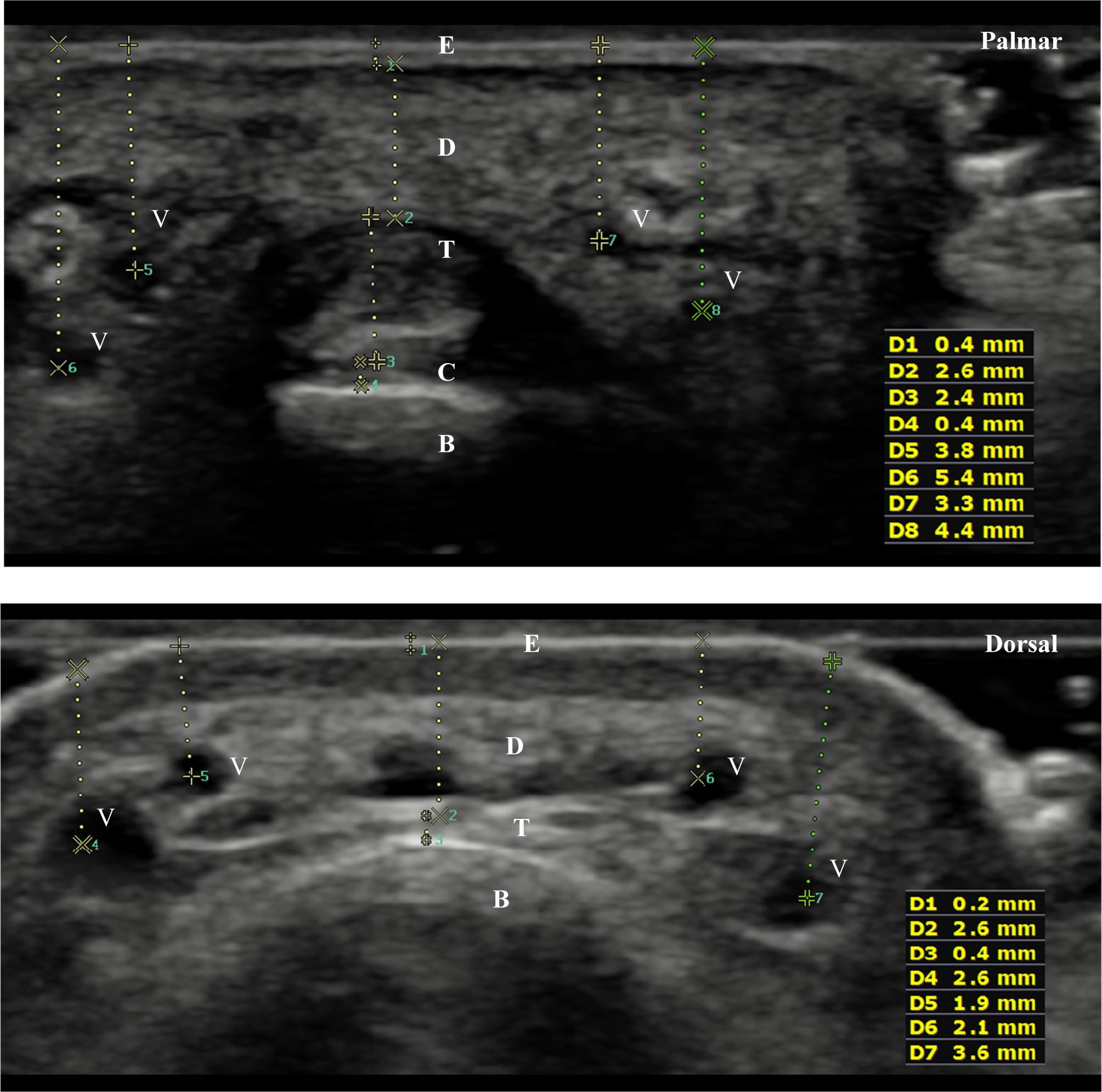}
\caption{Transverse orientation of the probe allowed us to image both the thickness of the entities and the distribution of vasculature. The palmar and dorsal side of the finger were imaged to measure the complete the finger structure.}
\label{fig:US_Trans}
\end{figure}

\begin{figure*} [t]
\begin{center}
\includegraphics[width=0.9\columnwidth]{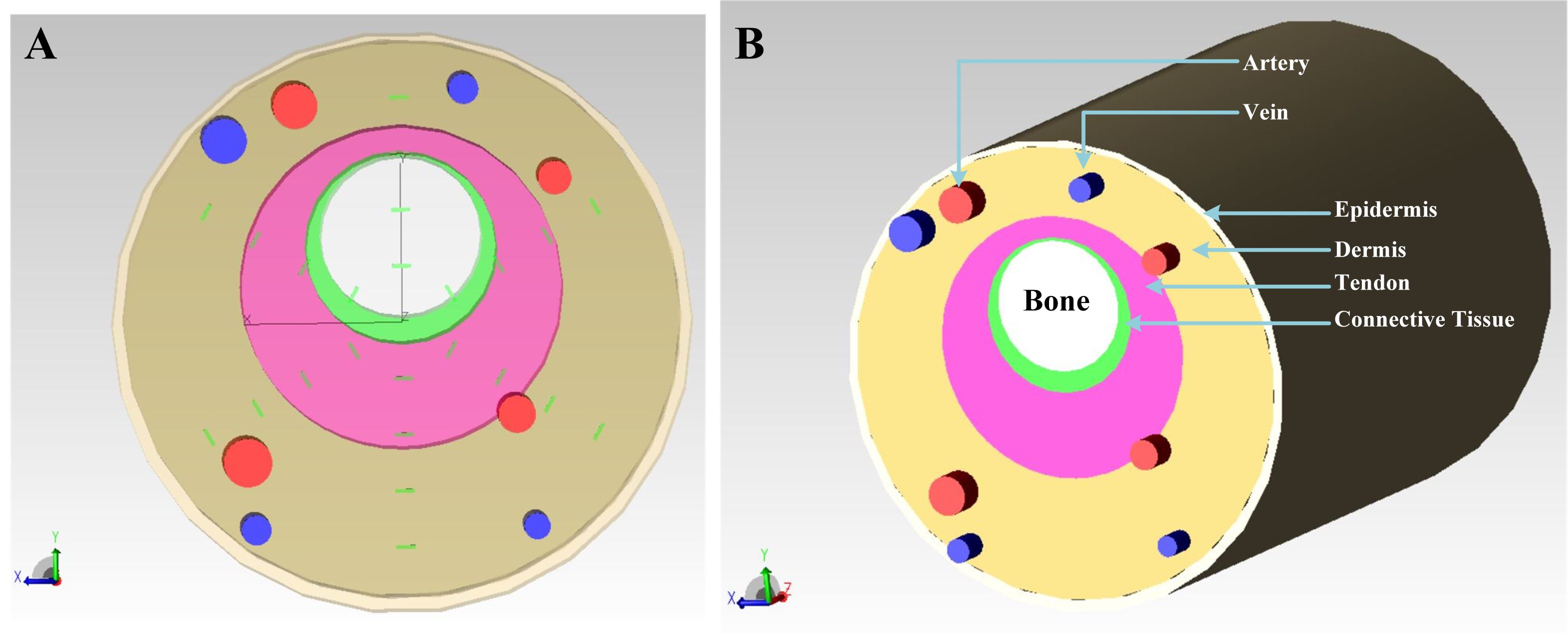}
\end{center}
\caption{The section view (A) of the optical model constructed from the ultrasound image data shows the placement of the anatomical entities and the distribution of the detectors. The detectors are placed at 60\textdegree\ intervals and at 1.5 mm, 3 mm, 4.5 mm and 6 mm radii from the centre. The rays are incident on the dorsal side of the finger, propagating in the -Z direction. The perspective view (B) shows the layers, with a length of 20 mm. The section "ends" of the finger absorb all incident light to disregard the rays that would propagate away from the tissue through these faces.}
\label{fig:TraceProUSMC_Model12}
\end{figure*}

For each participant, the middle phalanx of the left index finger was imaged. The palmar and dorsal sides of the finger were imaged with longitudinal (along the direction of the finger, figure \ref{fig:US_Long}) and transverse (perpendicular to the finger, figure \ref{fig:US_Trans}) orientations. In the images, we can identify and measure the thin echogenic epidermis, the more hypoechoic dermis, the flexor tendon and a thin connective tissue layer. We tabulated the geometry for each participant in Table \ref{tab:DepthThickUS}. The medial and lateral digital vessels were also identified where possible and the perpendicular distance from the skin surface to the centre of the vessel was measured. These measures were used to integrate the vasculature distribution in the optical model, as illustrated in Figure \ref{fig:TraceProUSMC_Model12}.

\subsection{Optical Properties}
As mentioned previously, the optical properties and their adaptation from literature is a key step. The optical properties required for both palmar and dorsal tissue (to the phalanx bone) for this study include the epidermis, dermis, tendons, connective tissue and bone. The epidermis and dermis layer properties are considered from previously published literature \cite{Jacques1991,Saidi1992,Altshuler2005,Simpson1998,Meglinski2002,Karsten2012,Tuchin2007_2}. The step of deciding the optical properties from literature is a key part of the methodology as previously, a 10-100 fold difference in optical properties have been reported \cite{Mignon2018}. The measurement methods, sample preparation and multiple other factors affect this quantification of optical properties, and therefore, the literature was reviewed critically before establishing the optical properties for each model. The subsequently presented models (Eq. \ref{eq:Anaemia_mua_skin}--\ref{eq:Anaemia_mua_dermis}) from the literature were used to calculate the optical properties. The $\mu_{a.epi}$ values for the epidermis are calculated by grouping the categories of the Fitzpatrick scale. Skin types I, II are combined and assigned a $f_{melanin}$ volume fraction of 0.038; type III, IV are set at $f_{melanin}=0.135$; and V, VI are set at $f_{melanin}=0.305$.

\begin{equation} \label{eq:Anaemia_mua_skin}
    \mu_{a.skin} = 7.84 \times 10^7 \times \lambda^{-3.255} \ [\lambda: nm] 
\end{equation}

\begin{equation} \label{eq:Anaemia_mua_melanin}
    \mu_{a.melanin} = 6.6 \times 10^{10} \times \lambda^{-3.3}  \ [\lambda: nm]
\end{equation}

\begin{equation} \label{eq:Anaemia_mua_epidermis}
    \mu_{a.epi} = (f_{melanin})(\mu_{a.melanin}) + (1-f_{melanin})(\mu_{a.skin})
\end{equation}

The dermal properties are calculated inclusive of blood perfusion with oxygen saturation set at 0.6, blood concentration at 0.04 and water fraction set at 0.6 \cite{Meglinski2002}. The factor, $\gamma$ is calculated as $Ht \times F_{RBC} \times F_{Hb}$. The factor $\gamma$ is used in eq. \ref{eq:Anaemia_mua_dermis}. The optical properties at the wavelengths simulated in the study are detailed in Table \ref{tab:OpticalProperties_TraceProVTS}.

\begin{equation} \label{eq:Anaemia_mua_dermis}
\begin{split}
    \mu_{a.der} &= (1-S) \ \gamma C_{blood} \ \mu_{a.Hb}(\lambda) + S \ \gamma C_{blood} \ \mu_{a.HbO2}(\lambda) \\ & + (1-\gamma C_{blood}) \ C_{H2O} \ \mu_{a.H2O}(\lambda) \\ & + (1-\gamma C_{blood})(1-C_{H2O})\mu_{a.skin}    
\end{split}
\end{equation}

\newpage

\begin{landscape}
\begin{table}[]
\centering
\caption{The thickness and depth measurements of the layers in the finger are used to construct the geometry of the optical models, with the inclusion of vasculature in the TracePro models. The vessels are not included in the semi-infinite slab models used in MCCL.}
\label{tab:DepthThickUS}
\begin{tabular}{lccccccccc}
\hline
\multicolumn{1}{c}{\multirow{2}{*}{\textbf{Participant}}} & \multicolumn{4}{c}{\textbf{Palmar}} & \multicolumn{3}{c}{\textbf{Dorsal}} & \multirow{2}{*}{\textbf{Bone}} & \multirow{2}{*}{\textbf{\begin{tabular}[c]{@{}c@{}}Fitzpatrick Scale\\ (Skin Type)\end{tabular}}} \\ \cline{2-8}
\multicolumn{1}{c}{} & \begin{tabular}[c]{@{}c@{}}Epidermis\\ \\ $d_{epi}$\end{tabular} & \begin{tabular}[c]{@{}c@{}}Dermis\\ \\ $d_{der}$\end{tabular} & \begin{tabular}[c]{@{}c@{}}Flexor \\ Profundus\\ $d_{ten}$\end{tabular} & \begin{tabular}[c]{@{}c@{}}Connective \\ Tissue \\ $d_{conn}$\end{tabular} & \begin{tabular}[c]{@{}c@{}}Epidermis\\ \\ $d_{epi}$\end{tabular} & \begin{tabular}[c]{@{}c@{}}Dermis \\ \\ $d_{der}$\end{tabular} & \begin{tabular}[c]{@{}c@{}}Extensor\\ Tendon\\ $d_{ten}$\end{tabular} &  &  \\ \hline
USP001 & 0.4 & 2.2 & 2.0 & 1.3 & 0.4 & 1.5 & 0.6 & 5 & I \\
USP002 & 0.4 & 3.8 & 1.9 & 0.9 & 0.3 & 1.8 & 0.5 & 6.5 & VI \\
USP003 & 0.4 & 4.1 & 2.1 & 0.4 & 0.3 & 2.0 & 0.4 & 7.3 & V \\
USP004 & 0.4 & 3.6 & 1.9 & 1.7 & 0.3 & 2.1 & 0.4 & 7.6 & IV \\
USP005 & 0.3 & 3.9 & 1.7 & 1.9 & 0.2 & 2.8 & 0.5 & 7 & II \\
USP006 & 0.4 & 3.1 & 1.8 & 1.7 & 0.2 & 2.1 & 0.4 & 7.4 & II \\
USP007 & 0.4 & 3.5 & 2.2 & 1.4 & 0.2 & 2.3 & 0.4 & 7.3 & IV \\
USP008 & 0.4 & 2.8 & 1.7 & 1.2 & 0.2 & 1.8 & 0.5 & 4.4 & IV \\
USP009 & 0.5 & 3.5 & 2.3 & 0.6 & 0.2 & 1.7 & 0.4 & 4.6 & I \\
USP010 & 0.5 & 3.6 & 1.9 & 1.7 & 0.3 & 2 & 0.6 & 4.1 & V \\
USP011 & 0.4 & 6.3 & 1.8 & 1.0 & 0.3 & 2.4 & 0.5 & 4.3 & VI \\
USP012 & 0.4 & 3.3 & 2.4 & 0.6 & 0.3 & 2.1 & 0.4 & 5.4 & V \\ \hline
\end{tabular}
\end{table}
\end{landscape}

\begin{figure}[h]
\centering
\includegraphics[width=0.45\columnwidth]{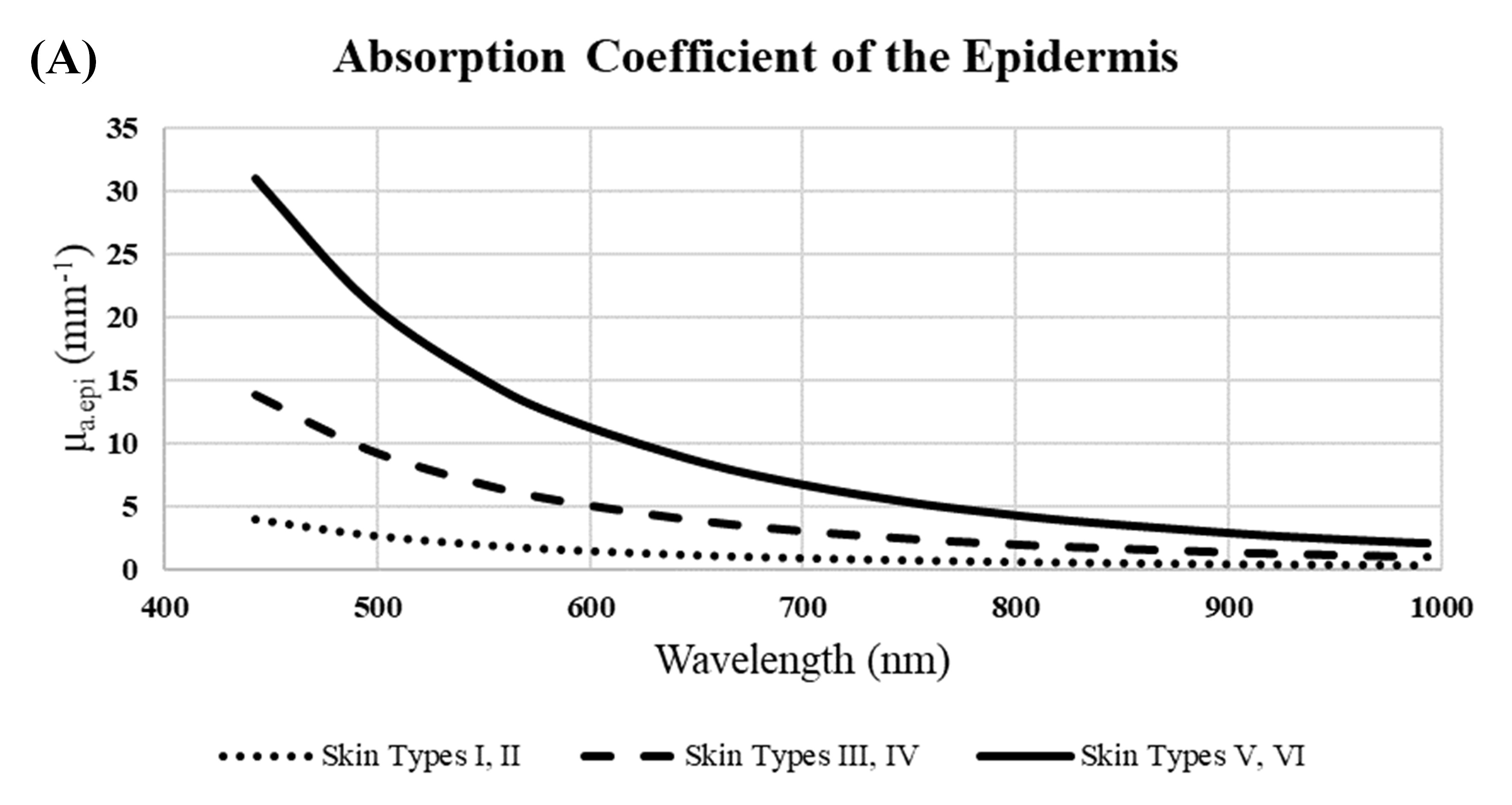}%
\includegraphics[width=0.45\columnwidth]{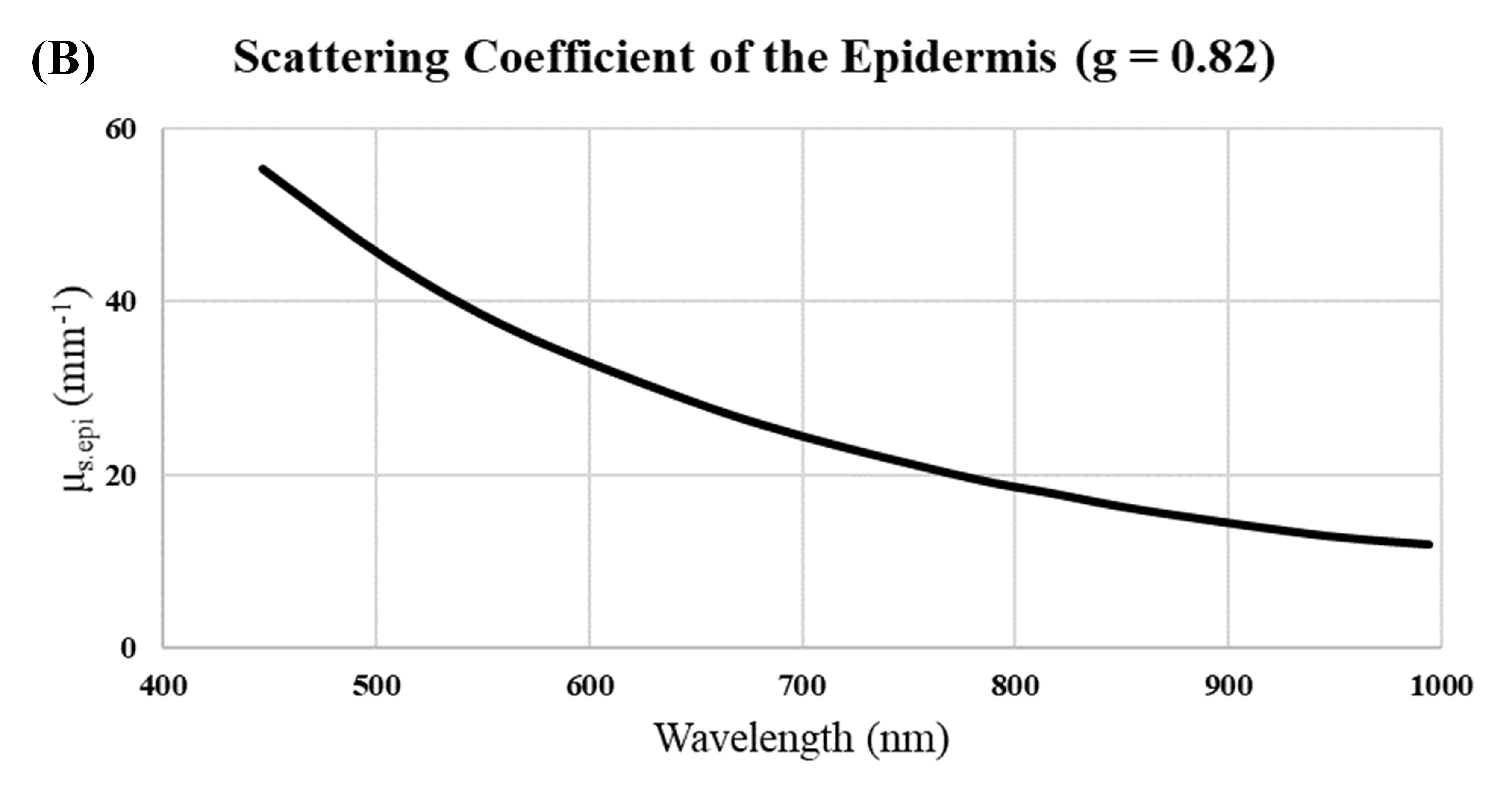}\\
\includegraphics[width=0.45\columnwidth]{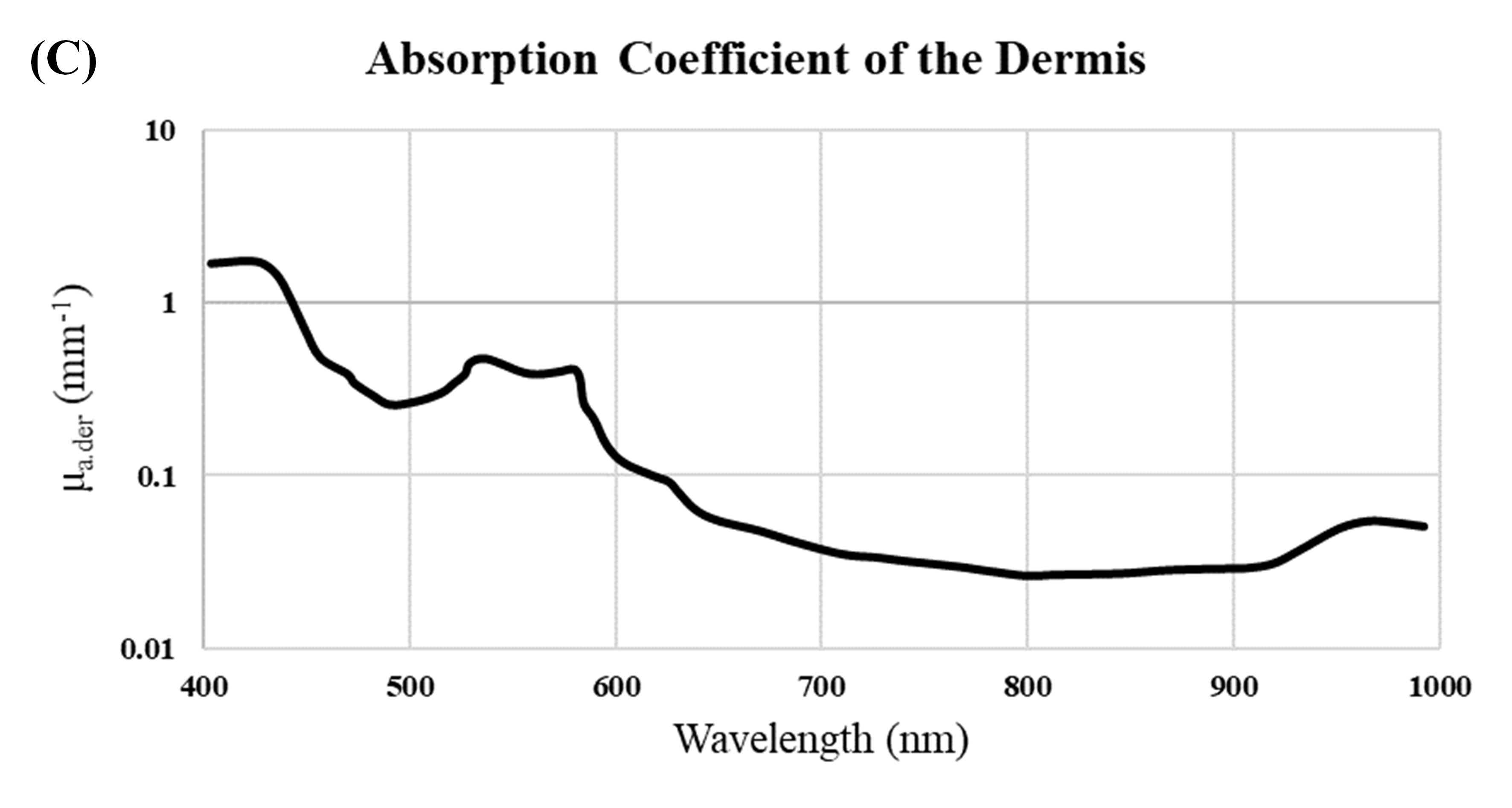}%
\includegraphics[width=0.45\columnwidth]{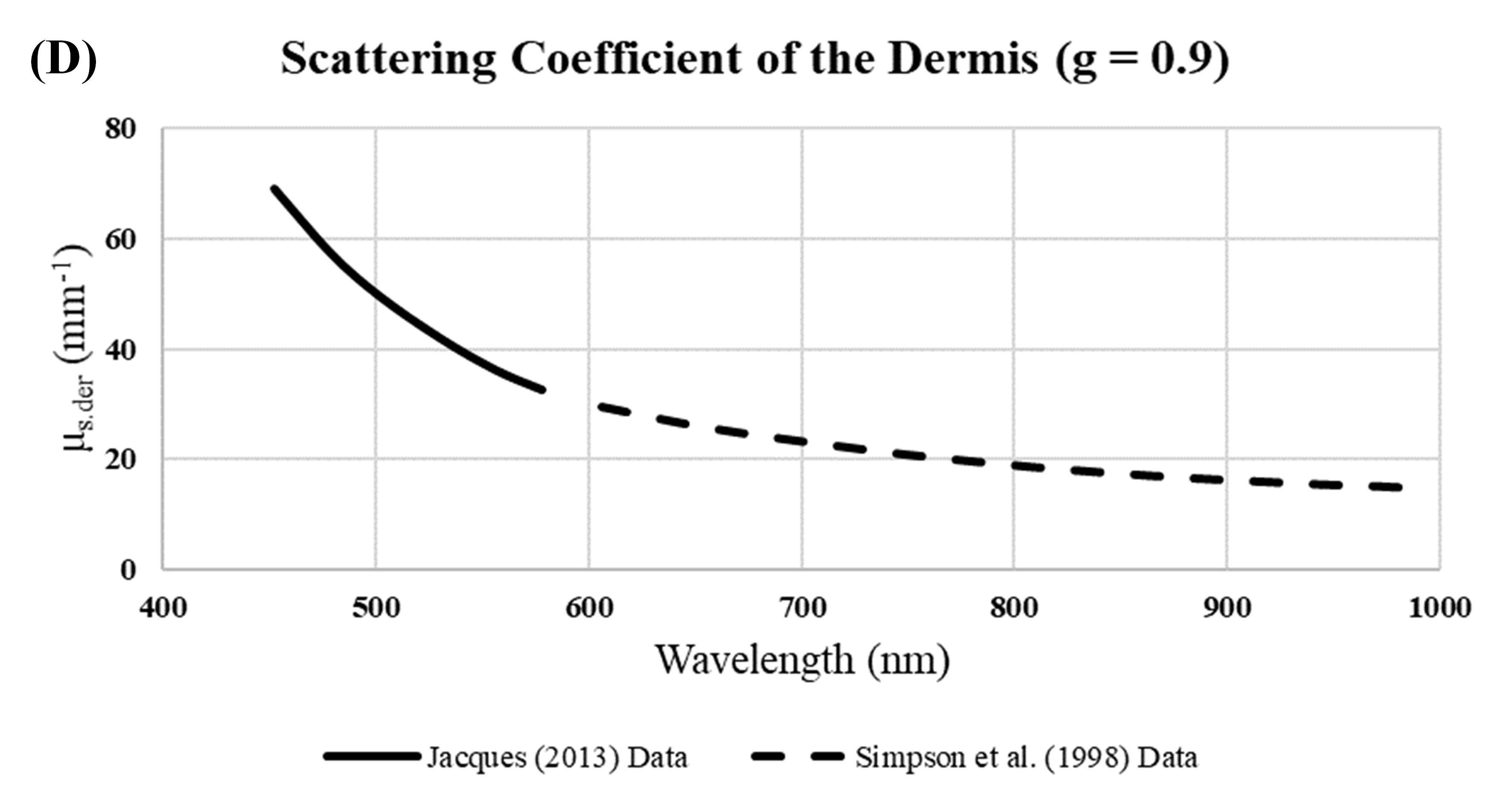}
\caption{The absorption properties of the epidermis (A) accommodate the varied distribution of melanin. The six skin types are modelled as three sub-groups, with associated $f_{melanin}$ volume fractions. The dermis considers water, oxygenated and deoxygenated haemoglobin in the perfused blood to quantify the absorption (C). Scattering is primarily due to the keratin and collagen fibres in the respective layers (B, D).}
\label{fig:USMC_EpidermisDermisProp}
\end{figure}

\begin{figure}[h]
\centering
\includegraphics[width=0.45\columnwidth]{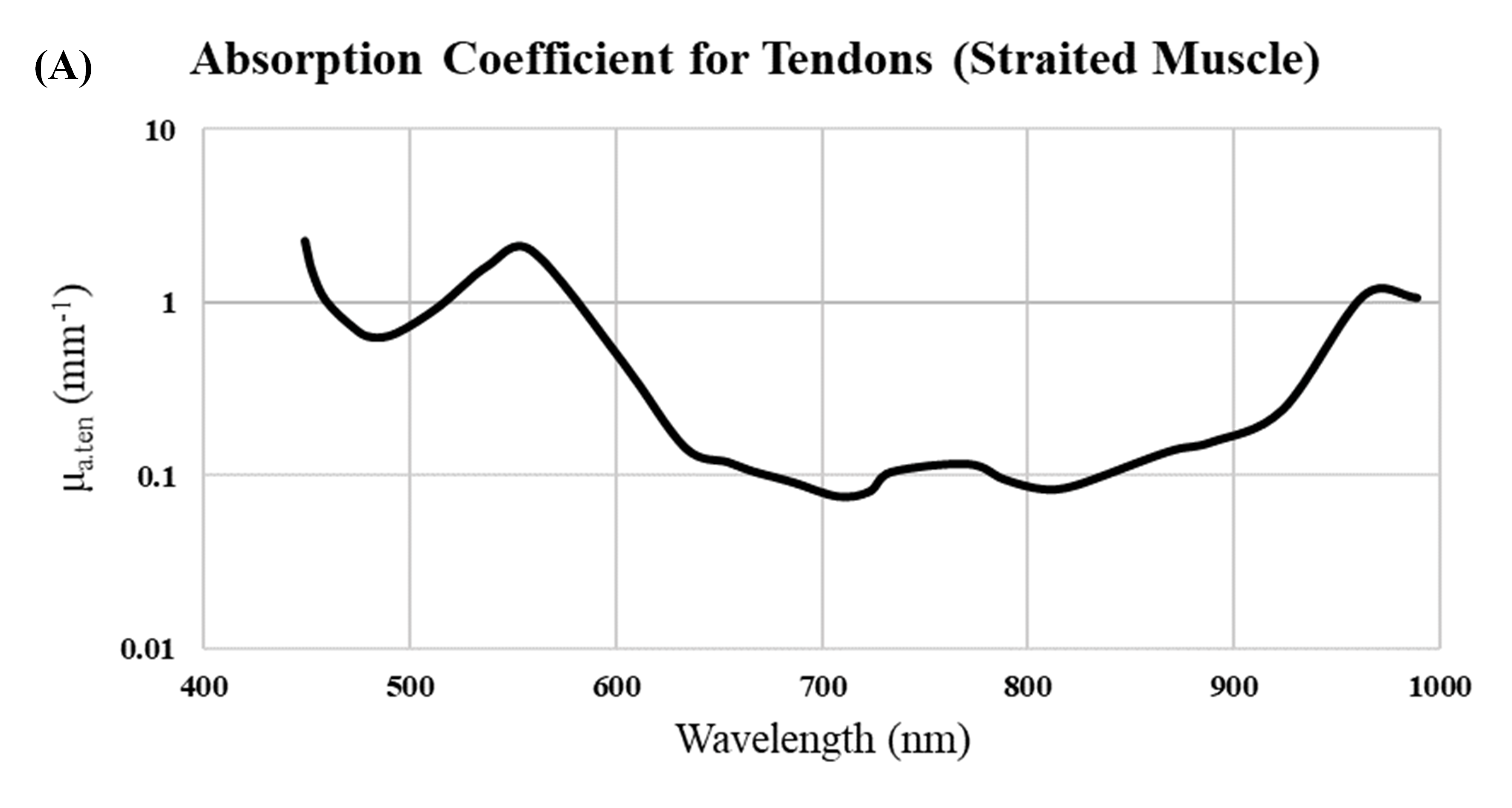}%
\includegraphics[width=0.45\columnwidth]{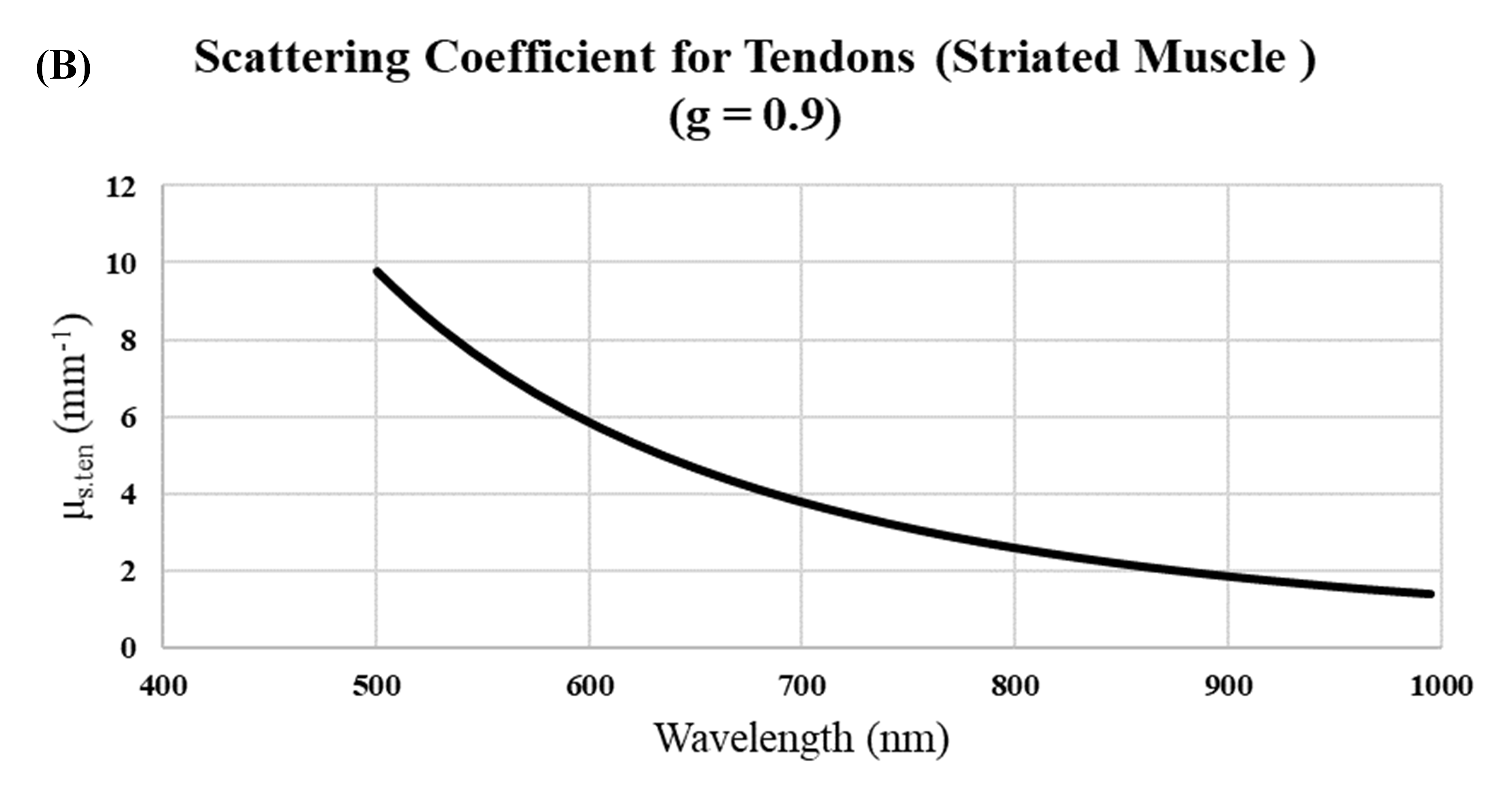}\\
\includegraphics[width=0.45\columnwidth]{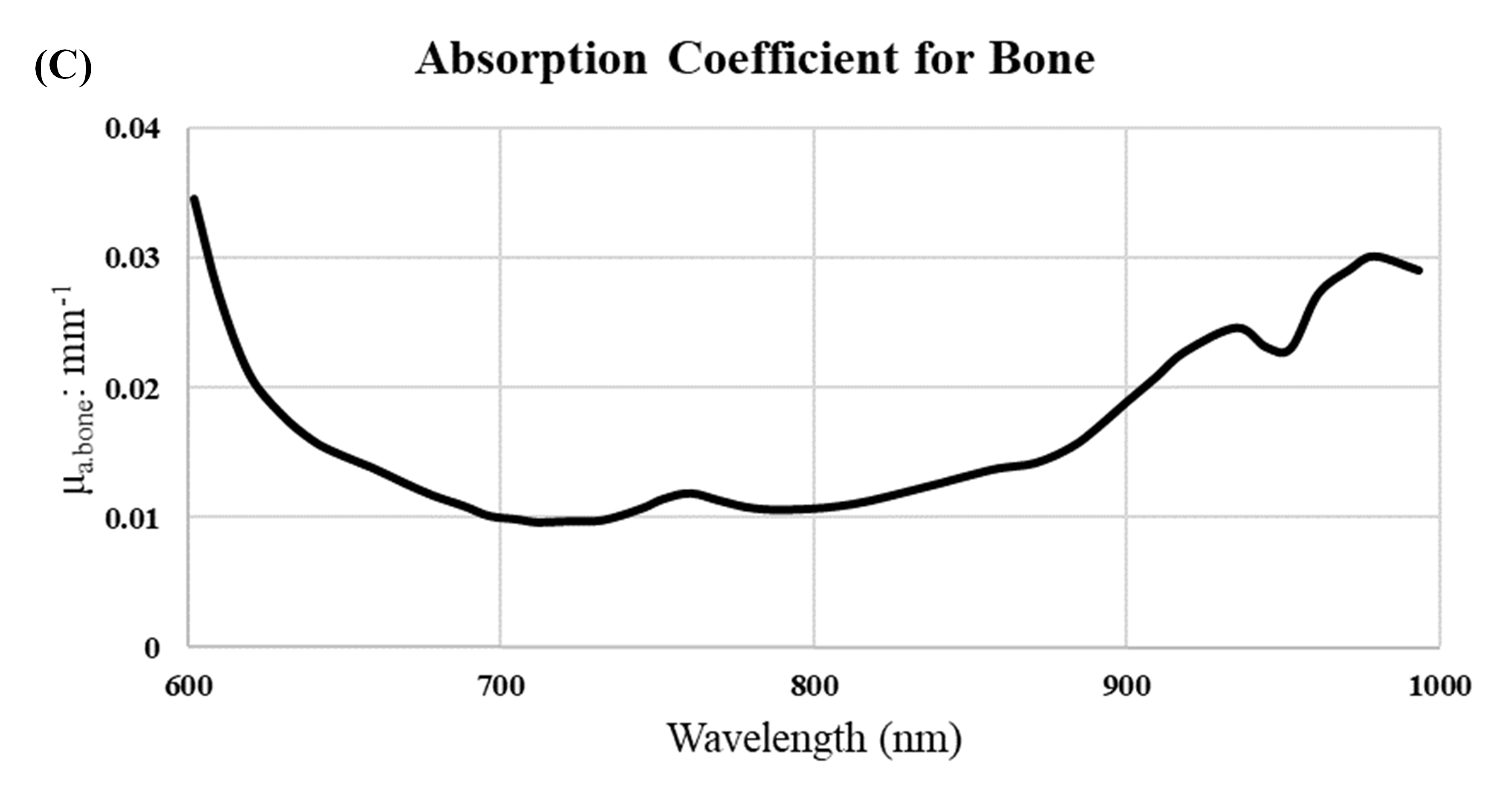}%
\includegraphics[width=0.45\columnwidth]{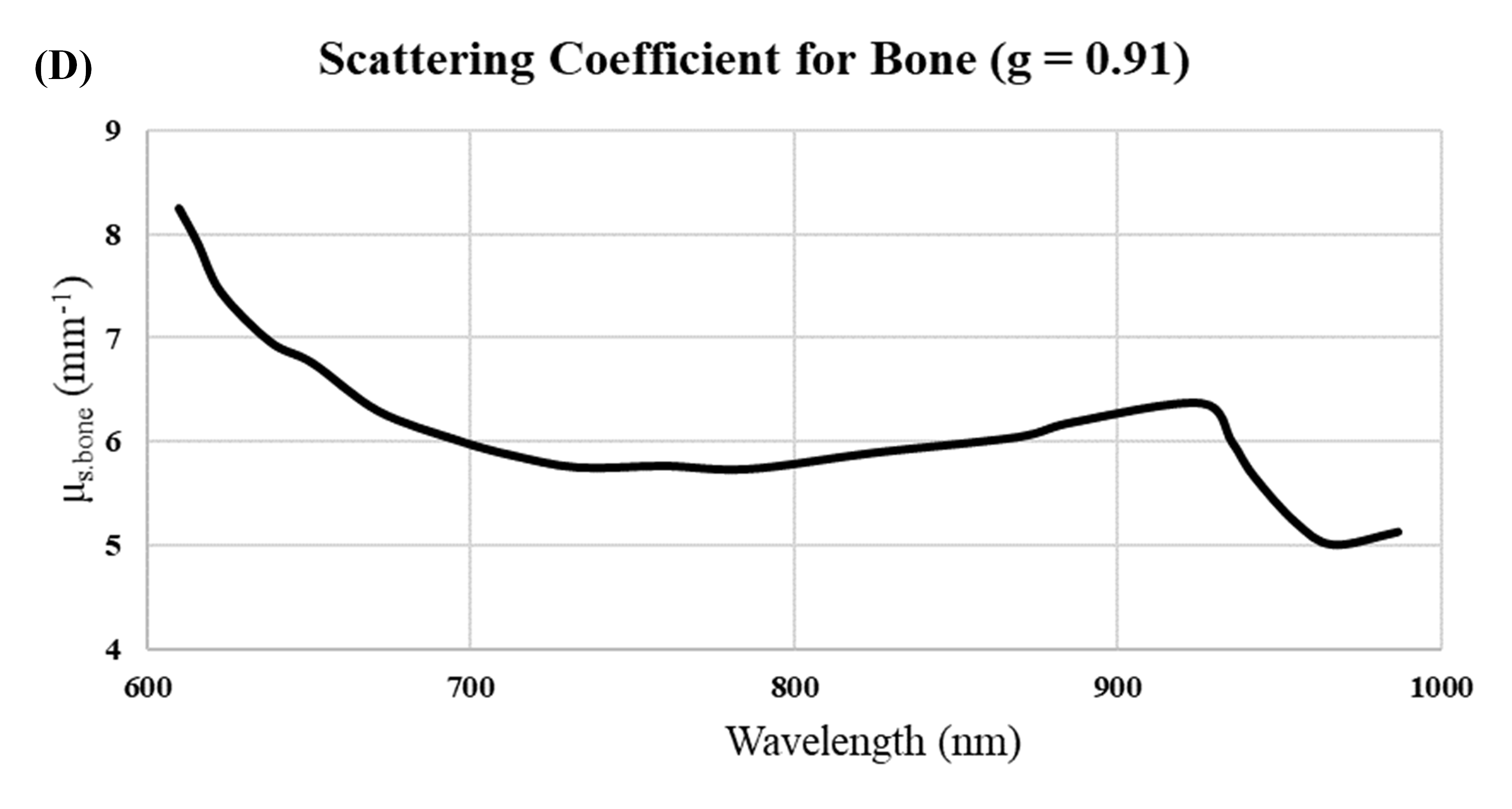}
\caption{The absorption and scattering coefficients for the tendon and bone in the finger show the influence of the tissue constituents. The studies measure the reduced scattering coefficient in a relatively simpler method and the anisotropy factor is used to further calculate the scattering coefficient.}
\label{fig:USMC_TendonBoneProp}
\end{figure}

Tendons in the human body are composed of fibrous muscle tissue and connect the muscles to the bone. The muscles themselves share properties similar to tendons. Alexandrakis \textit{et al.} \cite{Alexandrakis2005} studied the muscle as a part of addressing the need for a system combining tomographic bioluminescence imaging with optical-PET. The improved models were adopted and optical properties were calculated. In eq. \ref{eq:USMC_Mua}, $\alpha=0.07$ and $\beta=0.5$ represent heuristic scaling factors for blood and water concentration in the tissue. The ratio of oxygenated to total haemoglobin ($x=(HbO2)/(HbO2+Hb)$) is also considered constant at 0.8 \cite{Alexandrakis2005}. The values of $a$ and $b$ in eq. \ref{eq:USMC_Musp} that characterise the scattering in the tendon tissue are adopted as $4.e7$ and 2.82 from Kienle \textit{et al.} \cite{Kienle1996}. Additionally, the anisotropy factor is also ascertained as $g=0.9$. The refractive index is adopted as $\eta=1.37$ \cite{Tuchin2007_2}. 

\begin{equation} \label{eq:USMC_Mua}
\begin{split}
\mu_{a.ten}(\lambda) &= \alpha (x \ \mu_{a.Hb}(\lambda) + (1-x) \ \mu_{a.HbO2}(\lambda)) \\ & + \beta \ \mu_{a.H2O}(\lambda)
\end{split}
\end{equation}

\begin{equation} \label{eq:USMC_Musp}
\mu'_{s.ten}(\lambda) = a \times \lambda^{-b} 
\end{equation}

Using the above models, the absorption and scattering coefficients of the tendon were calculated (Figure \ref{fig:USMC_TendonBoneProp} A, B). The absorption coefficient shows the contributions of haemoglobin (peak between 500 nm and 600 nm) and water absorption (in the near-infrared region). The optical properties at relevant wavelengths for this study are given in Table \ref{tab:OpticalProperties_TraceProVTS}. 

\begin{table}[t]
\centering
\caption{The optical properties ($\mu_a$, $\mu_s$, $g$, $\eta$, $\mu'_s$) for the blood vessels quantify the contributions of the vessel, haemoglobin and oxygen saturation \cite{Tuchin2007_2}. The properties are specified in the respective spectrum regions as the exact properties for the wavelengths used in this study are unavailable. These properties are used in the TracePro simulation but not incorporated in the semi-infinite slab models used in MCCL (VTS).}
\label{tab:OpticalProperties_Vasculature_TracePro}
\begin{tabular}{ccccccc}
\hline
\textbf{\begin{tabular}[c]{@{}c@{}}Blood\\ Vessel\end{tabular}} & \textbf{\begin{tabular}[c]{@{}c@{}}Spectral\\ Range\end{tabular}} & \textbf{\begin{tabular}[c]{@{}c@{}}$\mu_a$\\ (mm$^{-1}$)\end{tabular}} & \textbf{\begin{tabular}[c]{@{}c@{}}$\mu_s$\\ (mm$^{-1}$)\end{tabular}} & \textbf{$g$} & \textbf{$\eta$} & \textbf{\begin{tabular}[c]{@{}c@{}}$\mu'_s$\\ (mm$^{-1}$)\end{tabular}} \\ \hline
\multirow{2}{*}{Artery} & Red & 0.13 & 124.6 & \multirow{4}{*}{0.9} & \multirow{2}{*}{1.38} & 6.11 \\
 & NIR & 0.28 & 50.5 &  &  & 3.84 \\
\multirow{2}{*}{Vein} & Red & 1.43 & 366 &  & \multirow{2}{*}{1.4} & 8.9 \\
 & NIR & 1.55 & 282 &  &  & 7.9 \\ \hline
\end{tabular}
\end{table}

The absorption and reduced scattering coefficients of the bone were measured using time-resolved diffuse optical spectroscopy (TRS) \cite{KonugoluVenkataSekar2016} and adopted for this study. Anisotropy measurements are adopted as $g=0.9$ from a different study, accounting for tissue thickness and measuring the reflectance and transmission \cite{Ugnell1997}. The refractive index ($\eta=1.55$) is adopted from \textit{in vivo} studies \cite{KonugoluVenkataSekar2016,Ascenzi1959}. The absorption and scattering coefficients used in the study are given in figure \ref{fig:USMC_TendonBoneProp} C, D and the values used in this study for the relevant wavelengths are summarised in Table \ref{tab:OpticalProperties_TraceProVTS}. The choice of wavelengths in the red and near-infrared region was due to the greater depth penetration in comparison to shorter wavelengths. 

The scattering properties of the connective tissue in the palmar side of the finger was assigned the scattering properties of the tendon (striated muscle) due to the anatomical similarity between muscles and connective tissues \cite{Kienle1996}. The absorption properties were, however, adopted from literature where the contribution of fatty tissue is considered \cite{Tuchin2007_2}. Although no fat layers were distinctly identified by the ultrasound system within the scope of this study, its contribution is included by considering the relevant optical properties.

\begin{table*}[p]
\centering
\caption{The optical properties ($\mu_a$, $\mu_s$, $g$, $\eta$, $\mu'_s$) are calculated and compiled, specific to this research, from various sources. The importance of screening optical properties specific to a study has previously been stated. These properties are applicable for MC methods in TracePro and MCCL (VTS). The properties of the vasculature (Table \ref{tab:OpticalProperties_Vasculature_TracePro}) are used in the TracePro simulations.}
\label{tab:OpticalProperties_TraceProVTS}
\begin{tabular}{ccccccc}
\hline
\textbf{Tissue} & \textbf{$\lambda$} & \textbf{\begin{tabular}[c]{@{}c@{}}$\mu_a$\\ (mm$^{-1}$)\end{tabular}} & \textbf{\begin{tabular}[c]{@{}c@{}}$\mu_s$\\ (mm$^{-1}$)\end{tabular}} & \textbf{$g$} & \textbf{$\eta$} & \textbf{\begin{tabular}[c]{@{}c@{}}$\mu'_s$\\ (mm$^{-1}$)\end{tabular}} \\ \hline
\multirow{4}{*}{\begin{tabular}[c]{@{}c@{}}Epidermis\\ (I, II)\\ $f_{melanin}=0.038$\end{tabular}} & 650 & 1.09 & 29.27 & \multirow{4}{*}{0.82} & \multirow{4}{*}{1.34} & 5.27 \\
 & 750 & 0.7 & 19.84 &  &  & 3.57 \\
 & 830 & 0.5 & 17.93 &  &  & 3.23 \\
 & 975 & 0.3 & 12.39 &  &  & 2.23 \\ \hline
\multirow{4}{*}{\begin{tabular}[c]{@{}c@{}}Epidermis\\ (III, IV)\\ $f_{melanin}=0.135$\end{tabular}} & 650 & 3.8 & 29.27 & \multirow{4}{*}{0.82} & \multirow{4}{*}{1.34} & 5.27 \\
 & 750 & 2.41 & 19.84 &  &  & 3.57 \\
 & 830 & 1.72 & 17.93 &  &  & 3.23 \\
 & 975 & 1.01 & 12.39 &  &  & 2.23 \\ \hline
\multirow{4}{*}{\begin{tabular}[c]{@{}c@{}}Epidermis\\ (V, VI)\\ $f_{melanin}=0.305$\end{tabular}} & 650 & 8.55 & 29.27 & \multirow{4}{*}{0.82} & \multirow{4}{*}{1.34} & 5.27 \\
 & 750 & 5.42 & 19.84 &  &  & 3.57 \\
 & 830 & 3.85 & 17.93 &  &  & 3.23 \\
 & 975 & 2.26 & 12.39 &  &  & 2.23 \\ \hline
\multirow{4}{*}{Dermis} & 650 & 0.05 & 26.54 & \multirow{4}{*}{0.9} & \multirow{4}{*}{1.4} & 2.65 \\
 & 750 & 0.03 & 20.93 &  &  & 2.09 \\
 & 830 & 0.03 & 17.77 &  &  & 1.78 \\
 & 975 & 0.05 & 15.21 &  &  & 1.52 \\ \hline
\multirow{4}{*}{Tendon} & 650 & 0.12 & 4.67 & \multirow{4}{*}{0.9} & \multirow{4}{*}{1.37} & 0.47 \\
 & 750 & 0.11 & 3.06 &  &  & 0.31 \\
 & 830 & 0.14 & 2.35 &  &  & 0.24 \\
 & 975 & 1.09 & 1.47 &  &  & 0.15 \\ \hline
\multirow{4}{*}{Bone} & 650 & 0.015 & 6.75 & \multirow{4}{*}{0.91} & \multirow{4}{*}{1.55} & 0.61 \\
 & 750 & 0.011 & 5.76 &  &  & 0.52 \\
 & 830 & 0.012 & 5.89 &  &  & 0.53 \\
 & 975 & 0.03 & 5.07 &  &  & 0.46 \\ \hline
\multirow{4}{*}{\begin{tabular}[c]{@{}c@{}}Connective\\ Tissue\end{tabular}} & 650 & 0.36 & 4.67 & \multirow{4}{*}{0.91} & \multirow{4}{*}{1.455} & 0.42 \\
 & 750 & 0.36 & 3.06 &  &  & 0.28 \\
 & 830 & 0.15 & 2.35 &  &  & 0.21 \\
 & 975 & 0.01 & 1.47 &  &  & 0.13 \\ \hline
\end{tabular}
\end{table*}

The optical properties of oxygenated and deoxygenated haemoglobin are adopted for the blood flowing through the arteries and veins directly from Table 2.1, Tuchin (2002) \cite{Tuchin2007_2}. These optical properties include contributions from the plasma and the blood in the blood vessels. The sources for optical properties specific to arteries and veins are sparse, although a lot is understood and published regarding haemoglobin itself \cite{Prahl2012_data,Jacques2013,Bashkatov2011}. The adopted values for the blood vessels are given in Table \ref{tab:OpticalProperties_Vasculature_TracePro}. The MC methods for analysing these models based on ultrasound images are applied in TracePro and MCCL. In MCCL, all four wavelengths mentioned in Table \ref{tab:OpticalProperties_TraceProVTS} are simulated. Due to the significantly longer computational time in TracePro, only red and near-infrared wavelength was simulated. The complete set of results are available through the CORD repository \cite{Kallepalli_USMCData_2020}. 

\begin{figure*}[htp]
\centering
\includegraphics[width=0.45\columnwidth]{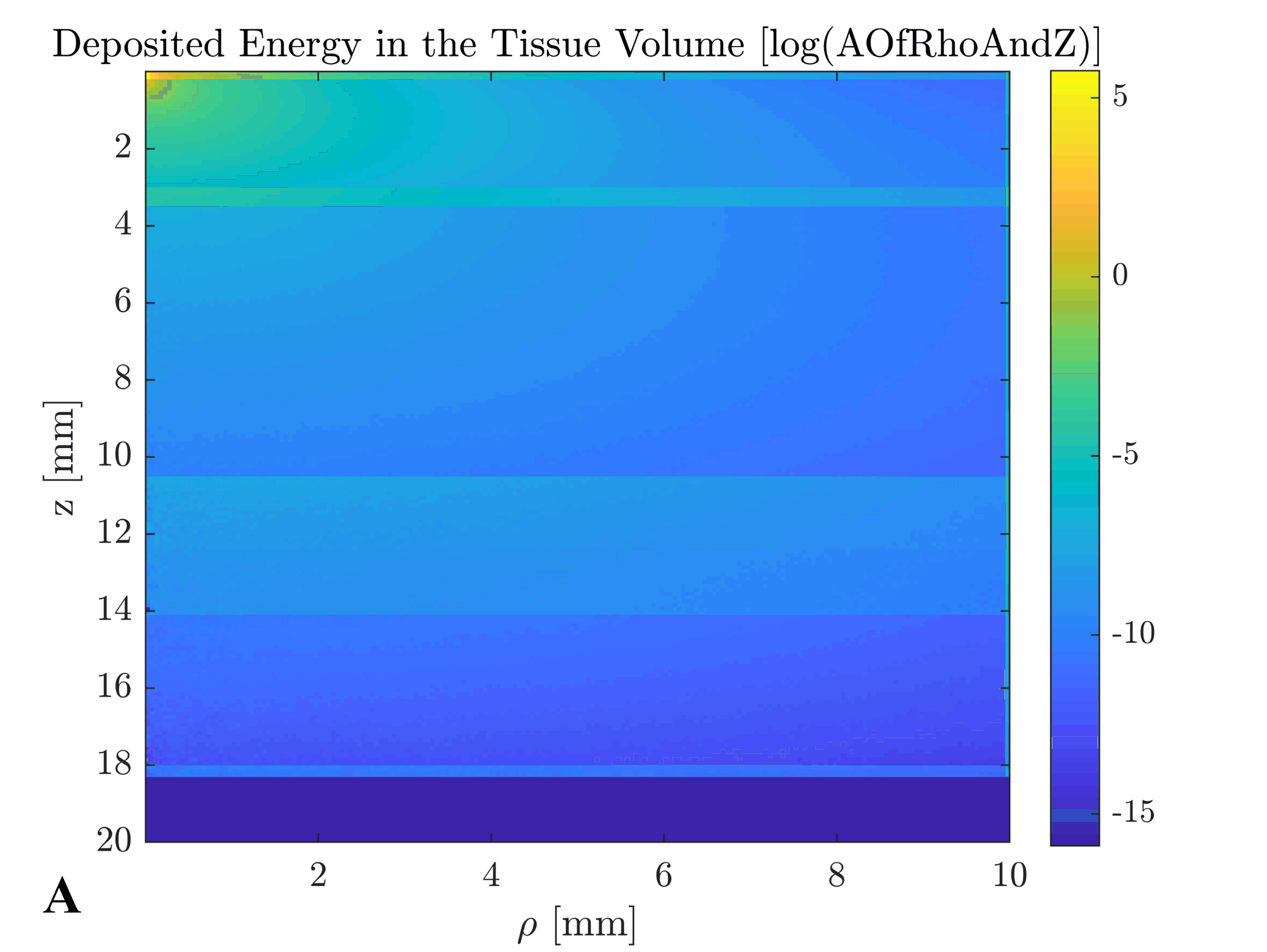}\
\includegraphics[width=0.45\columnwidth]{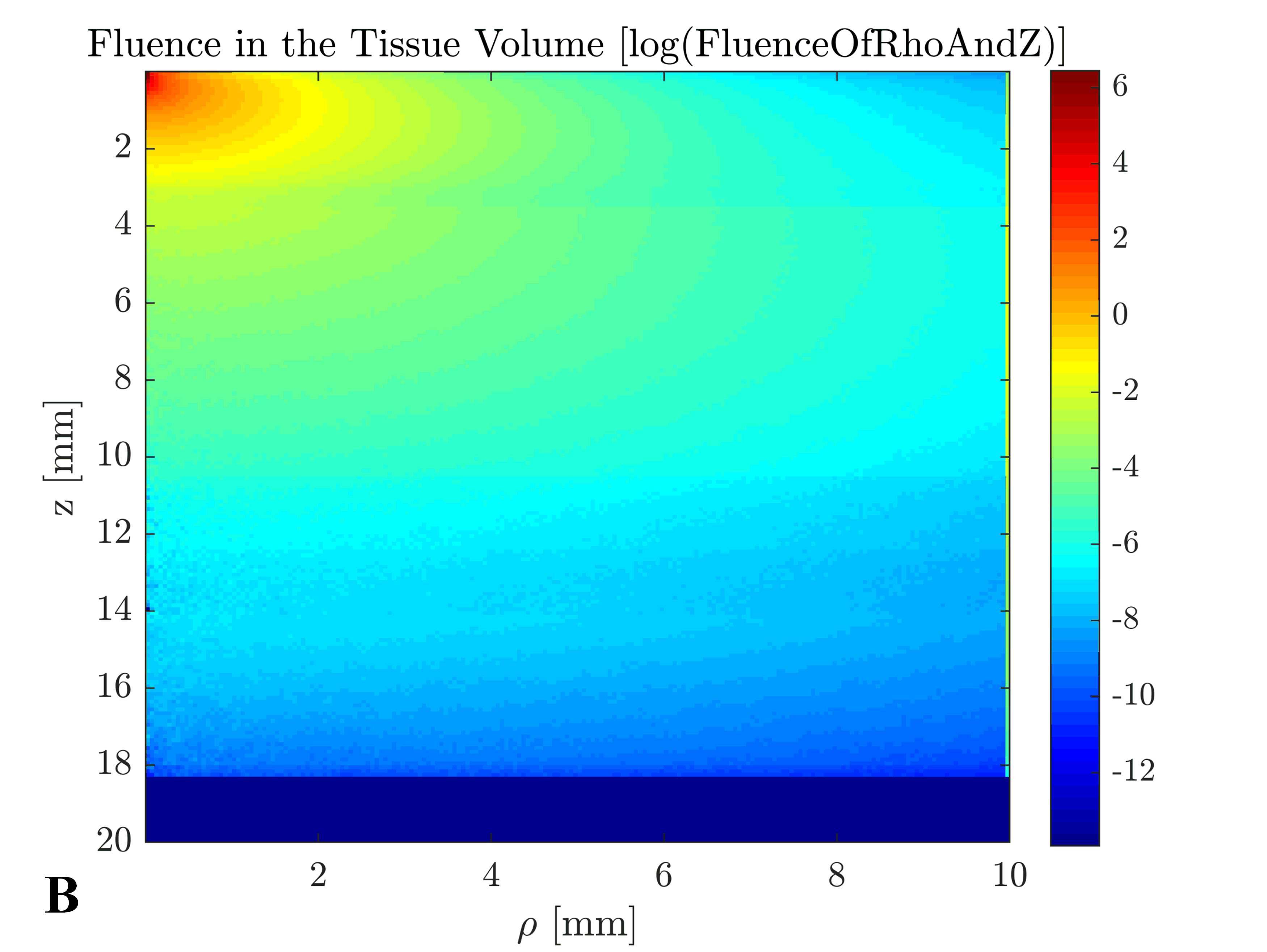}
\caption{The absorption (A) and fluence (B) in the tissue volume is seen for USP005 when using an 830 nm source. The absorption at each step within the layers clearly delineates the layers in the model, with the overall photon propagation illustrated with the fluence integrated over time.}
\label{fig:USMC_Result_AbsorptionFluence}
\end{figure*}

\subsection{Monte Carlo Methods}
Optical models require geometrical and optical properties to simulate the interaction with photons. The measurements from the ultrasound image data (Table \ref{tab:DepthThickUS}) are used to construct the geometry of the optical models, with some assumptions/approximations based on the simulation environment. Geometrical shapes can be constructed in 3D in the TracePro simulation environment. Therefore, the thickness and depths of the layers were adapted and the middle phalanx region of the finger was approximated to a cylinder (Figure \ref{fig:TraceProUSMC_Model12}). Additionally, the presence of blood vessels can be accommodated in TracePro, with the relevant optical properties. The layers of the finger are defined as semi-infinite slabs in MCCL, placed sequentially from the epidermis in the dorsal side of the finger to the palmar side (i.e. epidermis, dermis, extensor tendon, bone, connective tissue, flexor profundus, dermis and epidermis). 

While TracePro monitors and saves every path a ray takes within and beyond the boundaries of the tissue model, MCCL requires designated "detectors" that measure individual parameters related to quantifying photon interactions. The MCCL detectors measure the absorbed photons (AOfRhoAndZ) in the tissue volume as a function of cylindrical coordinates, $\rho$ (radius from the point of incidence) and depth, $z$. The fluence is measured and shows the propagation of photons function of $\rho$ and $z$. When integrated over time intervals, the FluenceOfRhoAndZAndTime detector is used, while FluenceOfRhoAndZ measures the fluence across the entire duration of the photon transport. Finally, the reflected energy (ROfRho) and transmitted energy (TOfRho) are measured as a function of $\rho$ on the bottom surface of the model. The workflow and details of using MCCl are given online (\url{https://github.com/VirtualPhotonics/VTS/wiki/Virtual-Tissue-Simulator}) and not repeated here for brevity. 

\section{Results and Discussion} \label{sec:res_Disc}
The methodology presented in this research images and obtains individual-specific anatomy for unique and accurate modelling of light transport in the tissue. The application of this method is site-independent as modelling of reflected and transmitted photons is possible. Therefore, may it be at the finger or the abdomen where the light cannot transmit through the tissue bulk, this methodology is relevant and applicable. 
\begin{figure*}[htp]
\centering
\includegraphics[width=0.33\columnwidth]{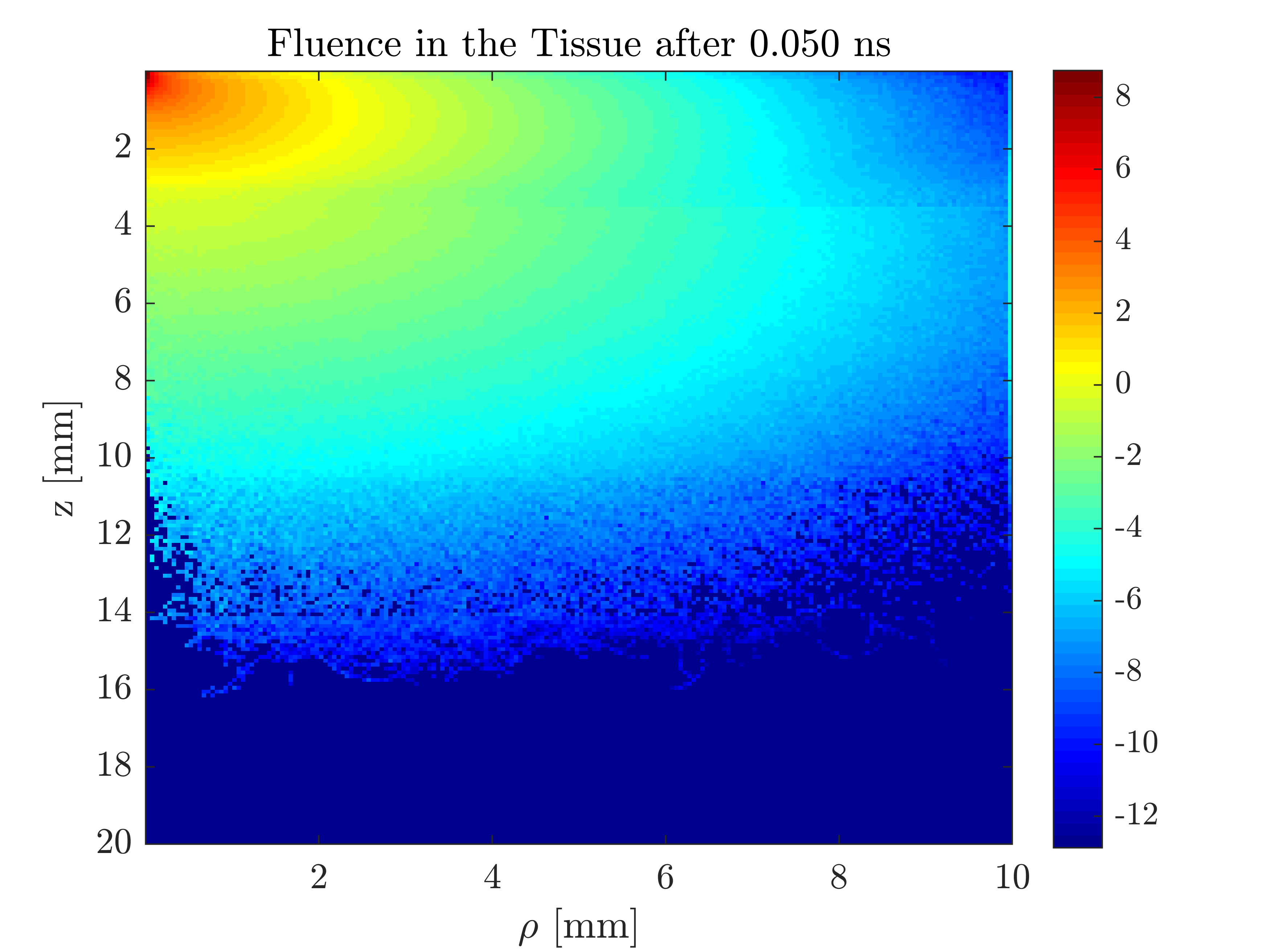}
\includegraphics[width=0.33\columnwidth]{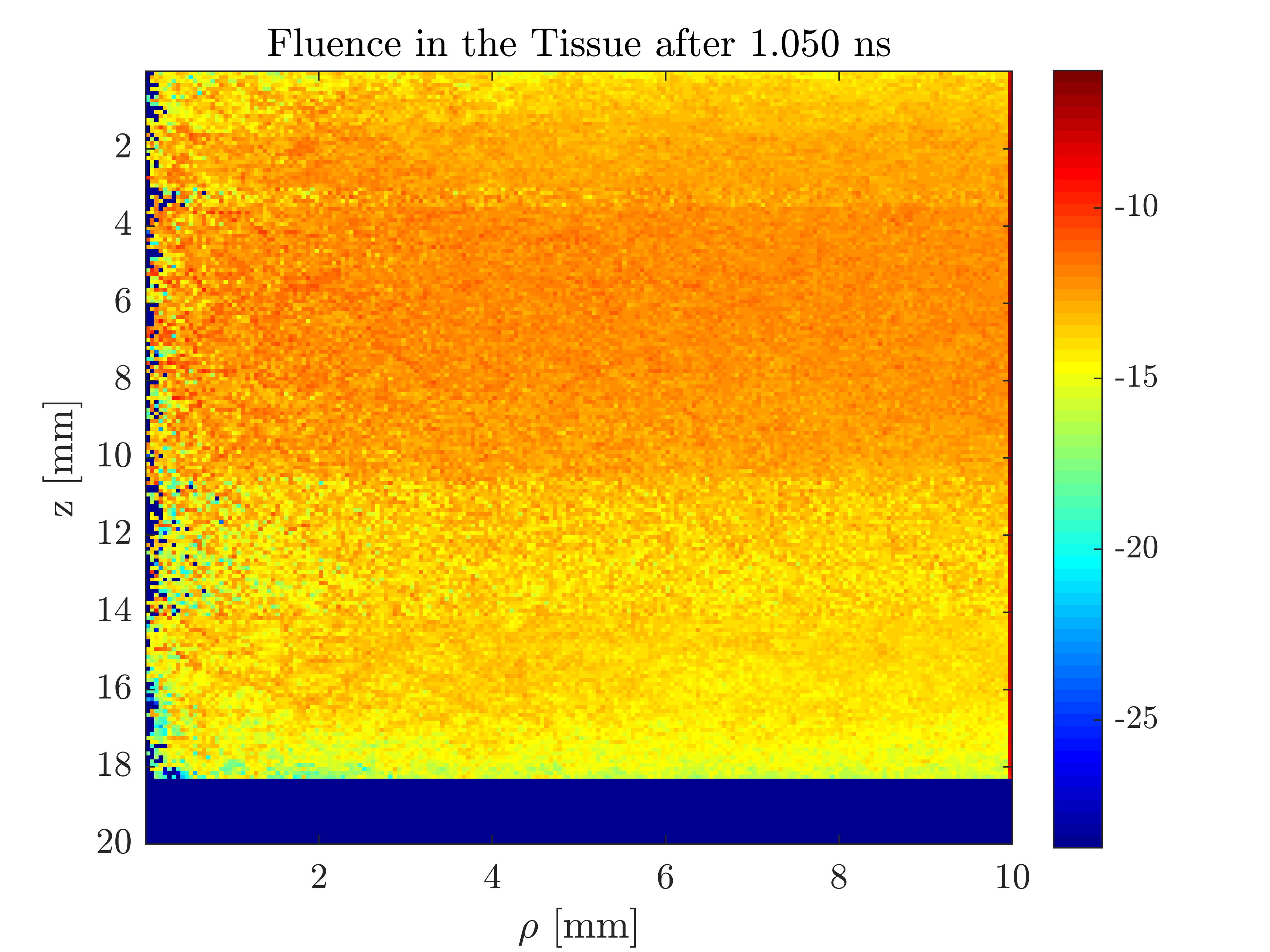}
\includegraphics[width=0.33\columnwidth]{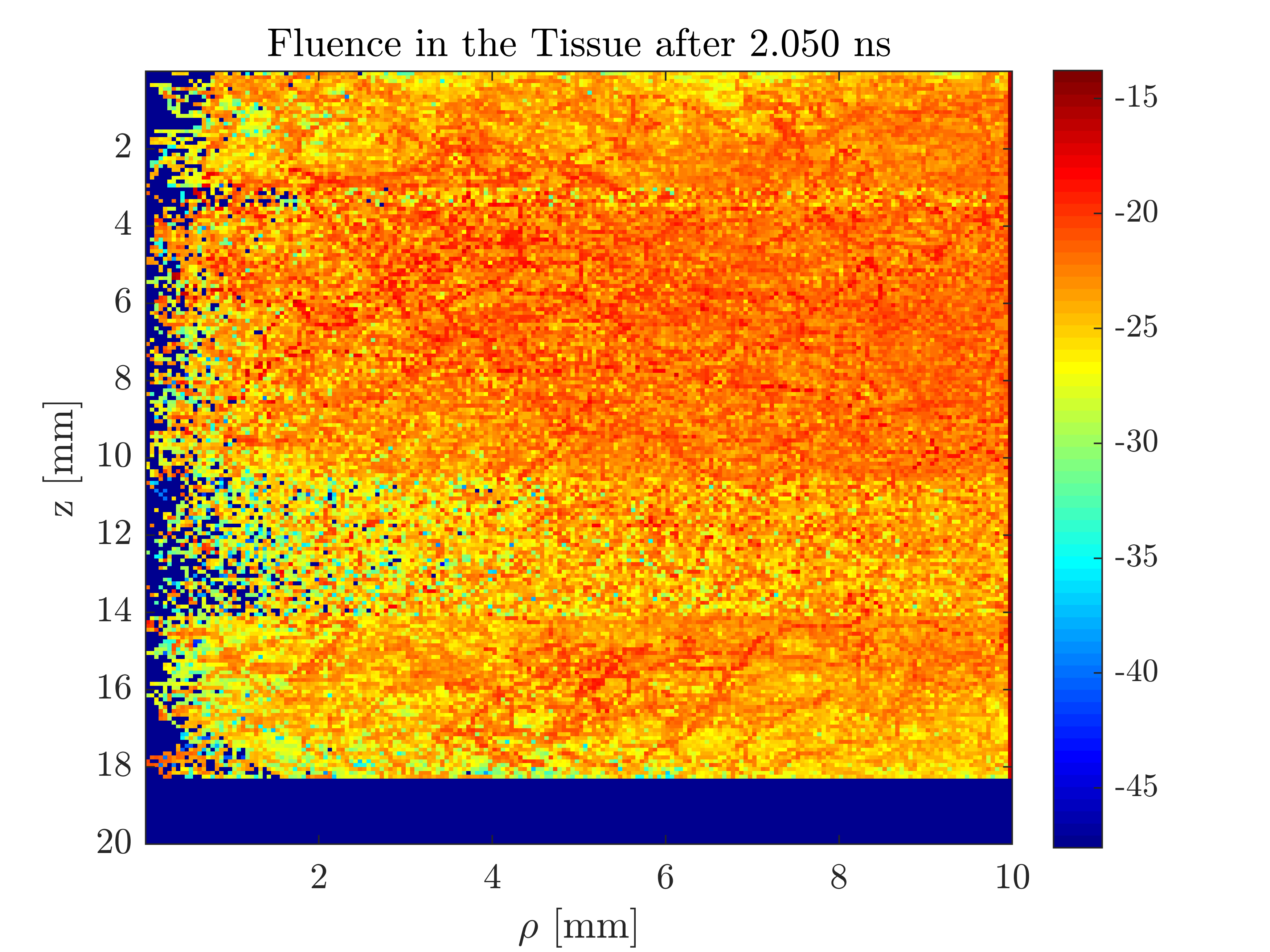}\\
\includegraphics[width=0.33\columnwidth]{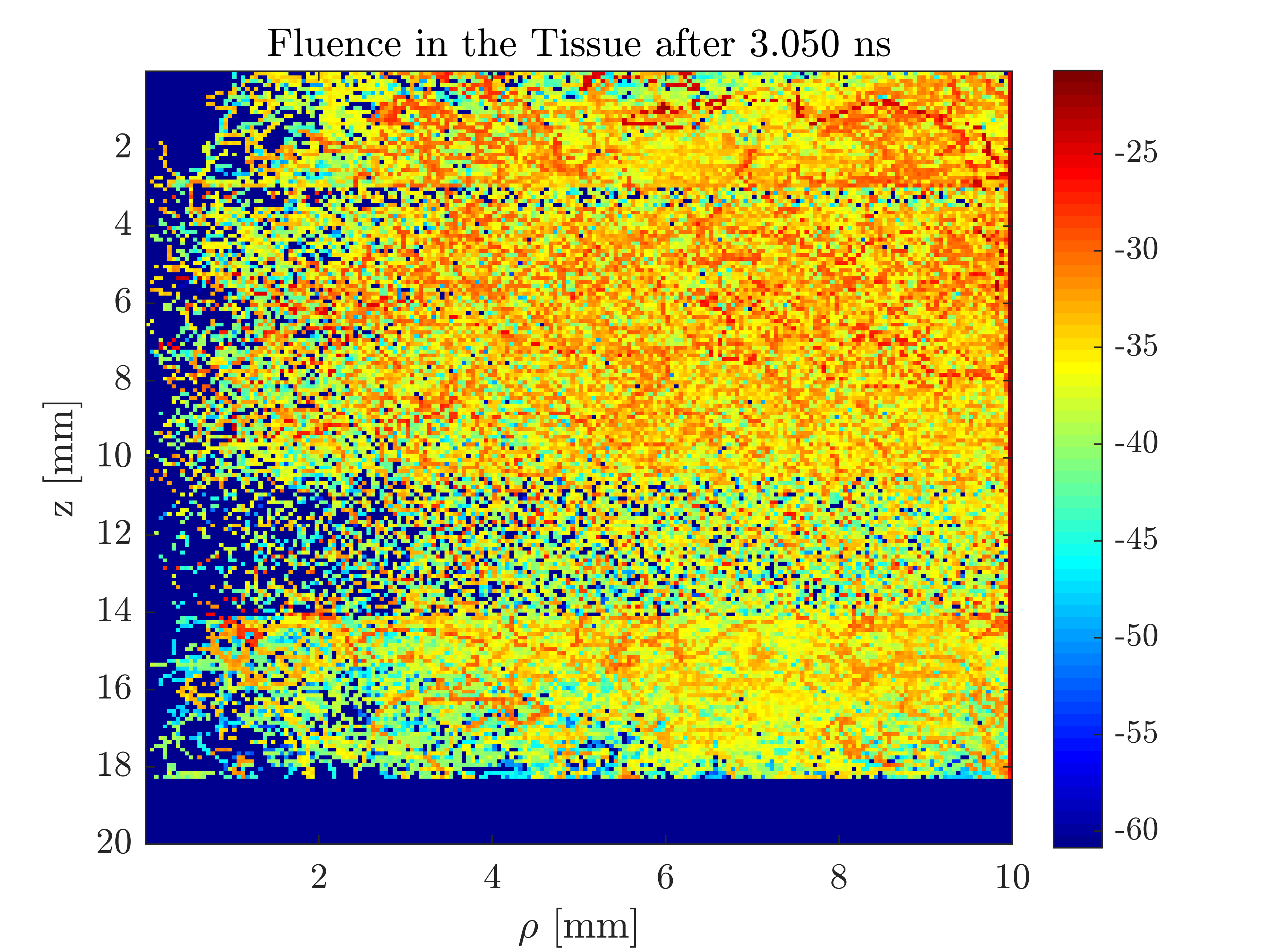}
\includegraphics[width=0.33\columnwidth]{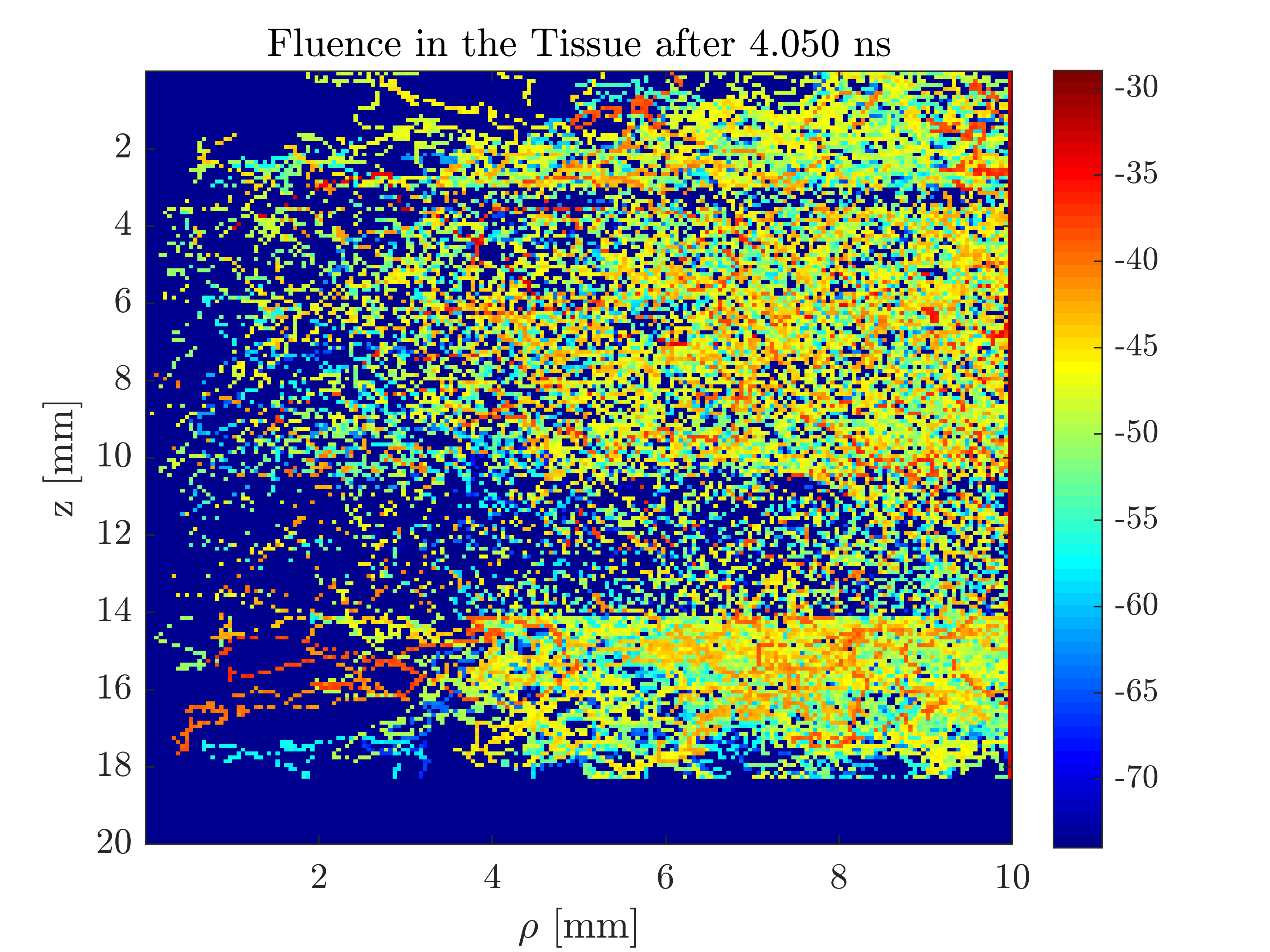}
\includegraphics[width=0.33\columnwidth]{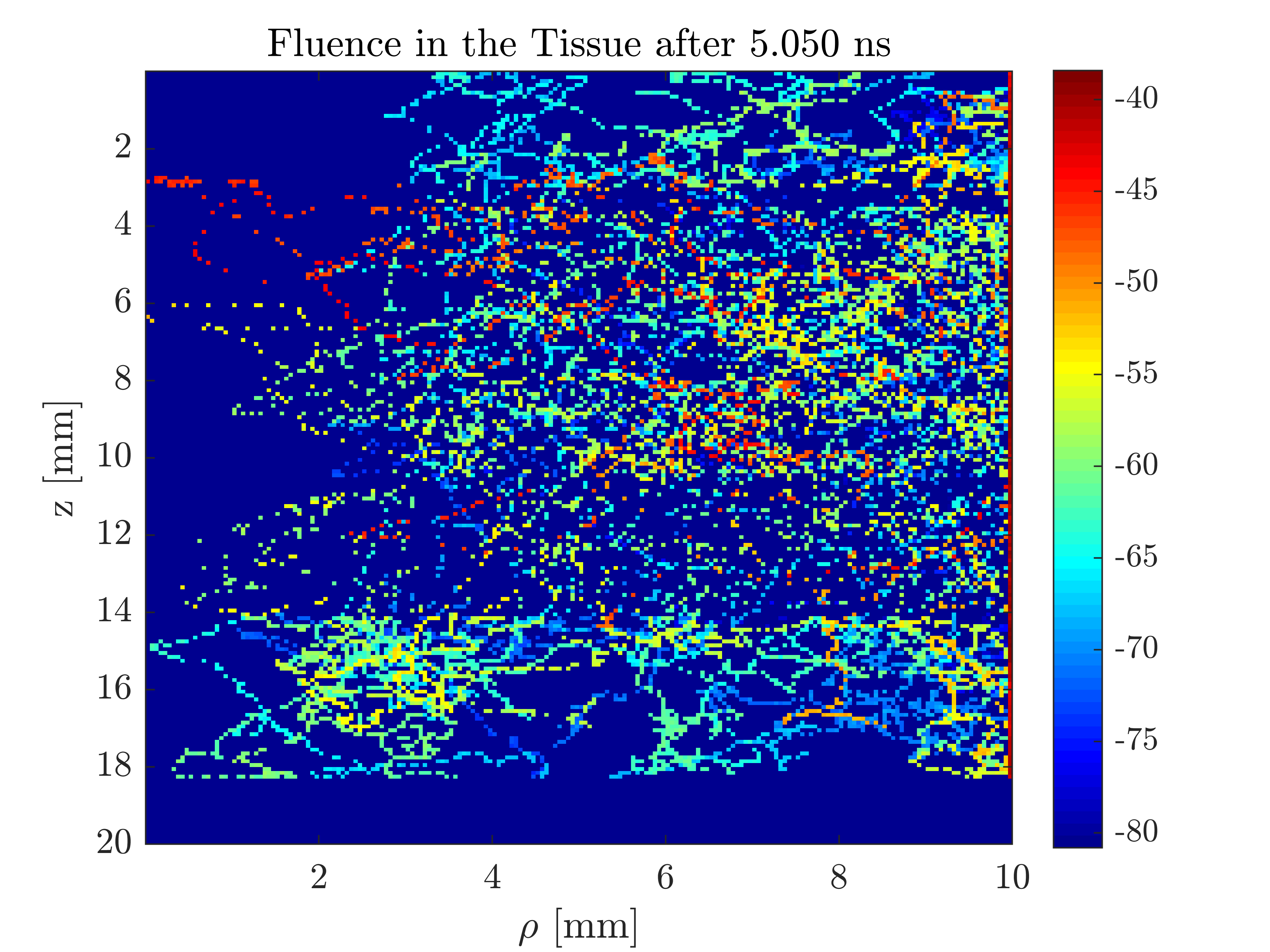}
\caption{The fluence of the photons at 830 nm wavelength in the tissue volume (mW/mm$^{-3}$) as a function of cylindrical coordinates ($\rho$: mm) and depth ($z$) is illustrated at different instances after emission from the sources for the optical model for participant USP005 (Table \ref{tab:DepthThickUS}).}
\label{fig:USMC_Result_Fluence6}
\end{figure*}

The amount of optical energy deposited at different depths can be inferred from Figure \ref{fig:USMC_Result_AbsorptionFluence} A. The propagation of photons (Figure \ref{fig:USMC_Result_AbsorptionFluence} B) illustrates the fluence integrated over time through the optical model. This optical model allows deciphering cancerous or anomalous tissue, and the amount of energy deposited at the site. This information can be used to decide on the power and wavelength of the therapeutic laser. Fewer photons propagate through the bulk of the entire tissue. Further, the fluence of the photons can be presented as a function of time (Figure \ref{fig:USMC_Result_Fluence6}), illustrating photon propagation at each time interval. This inference will allow evidence-based decision making with flexibility of deciding the wavelength, power of the laser and the duration of exposure in optical therapy. Within the context of this study, the skin-safe laser power and wavelengths do not require an in-simulation threshold to assess the damage to the tissue. However, this is easily included in the data analysis to highlight potential damage to tissue at sites with healthy tissue. This is beneficial in applications such as photo-dynamic therapy and photo-modulation. 

\begin{figure*}[htp]
\centering
\includegraphics[width=0.49\columnwidth]{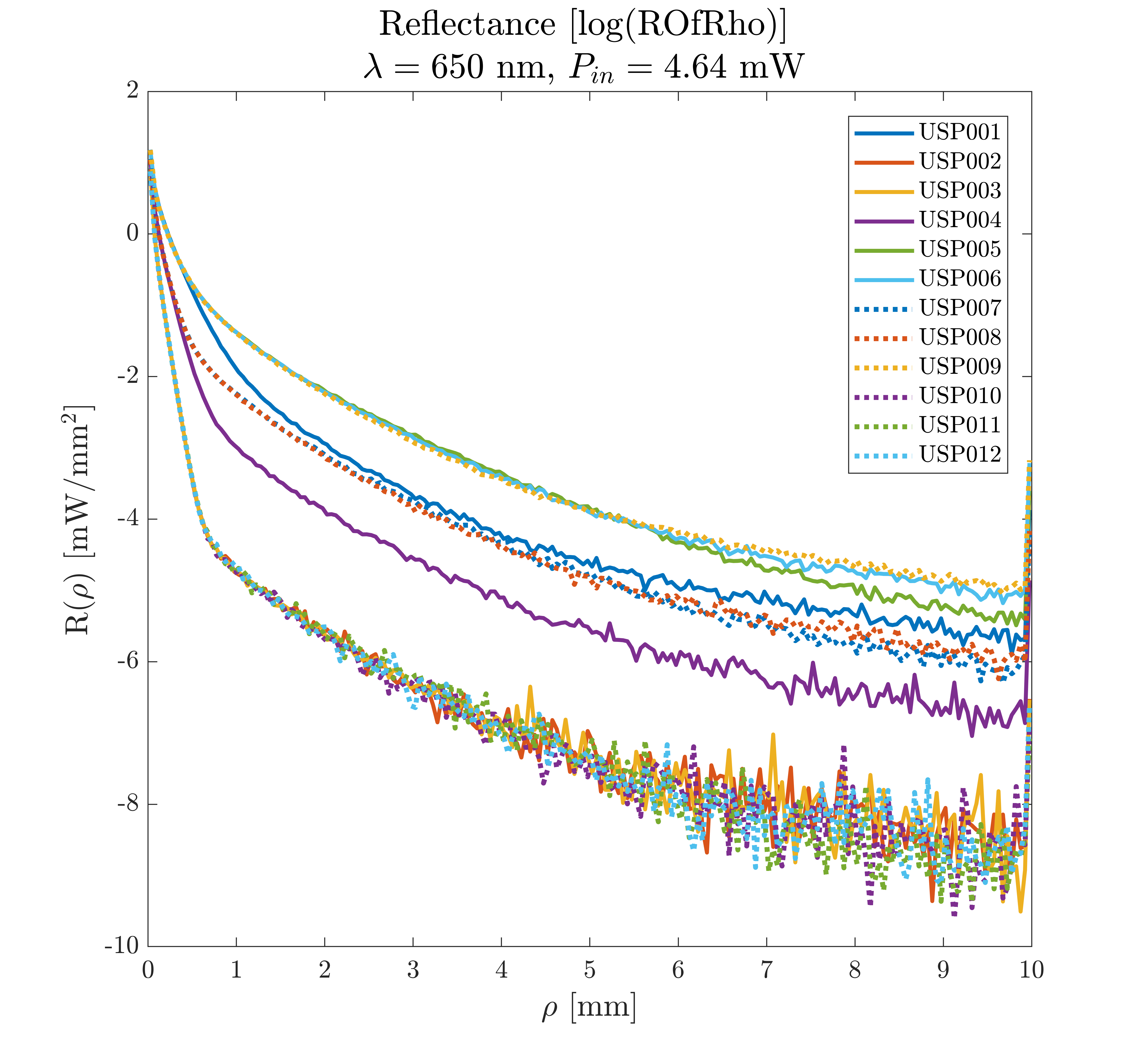}%
\includegraphics[width=0.49\columnwidth]{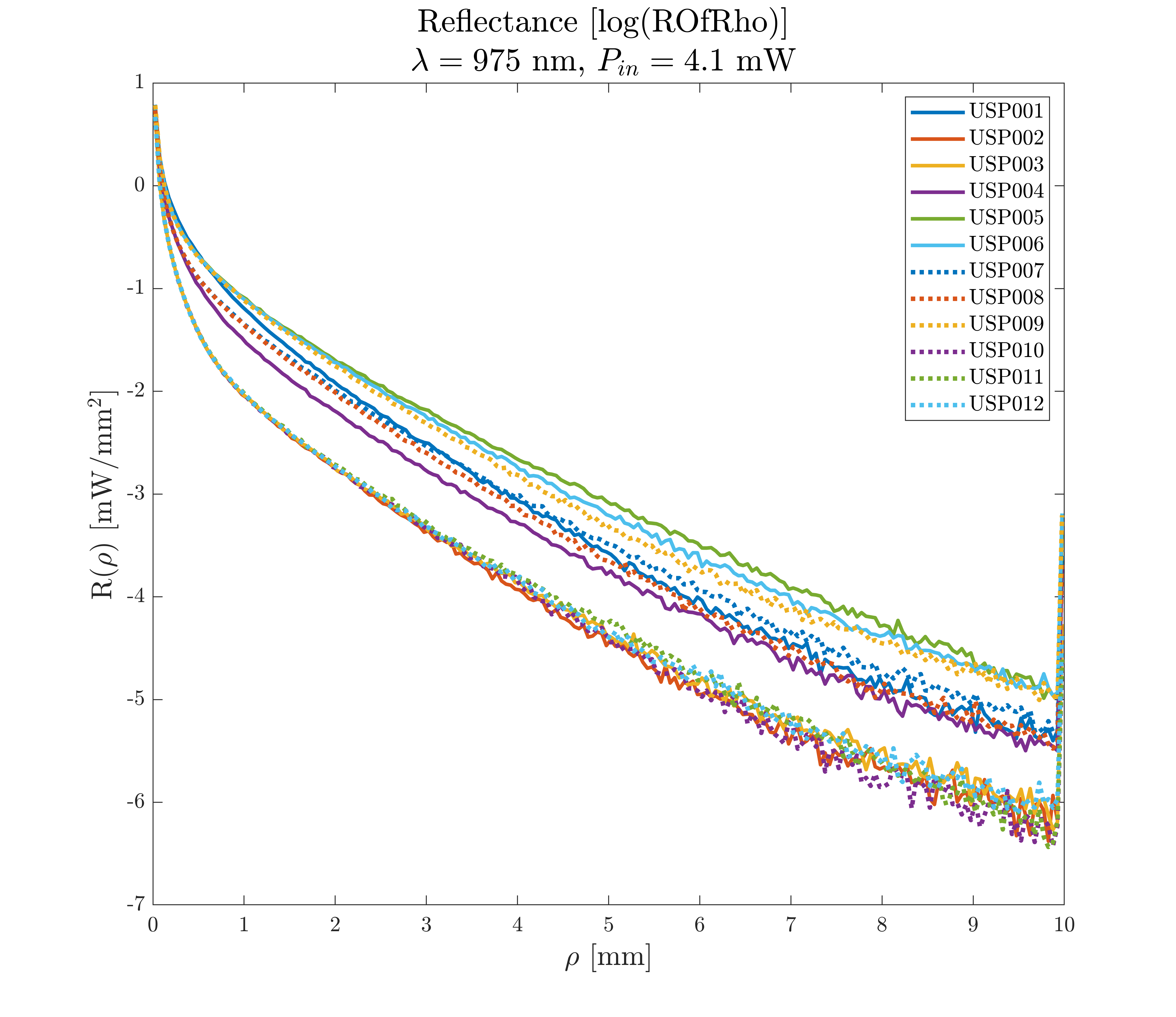}
\\
\includegraphics[width=0.49\columnwidth]{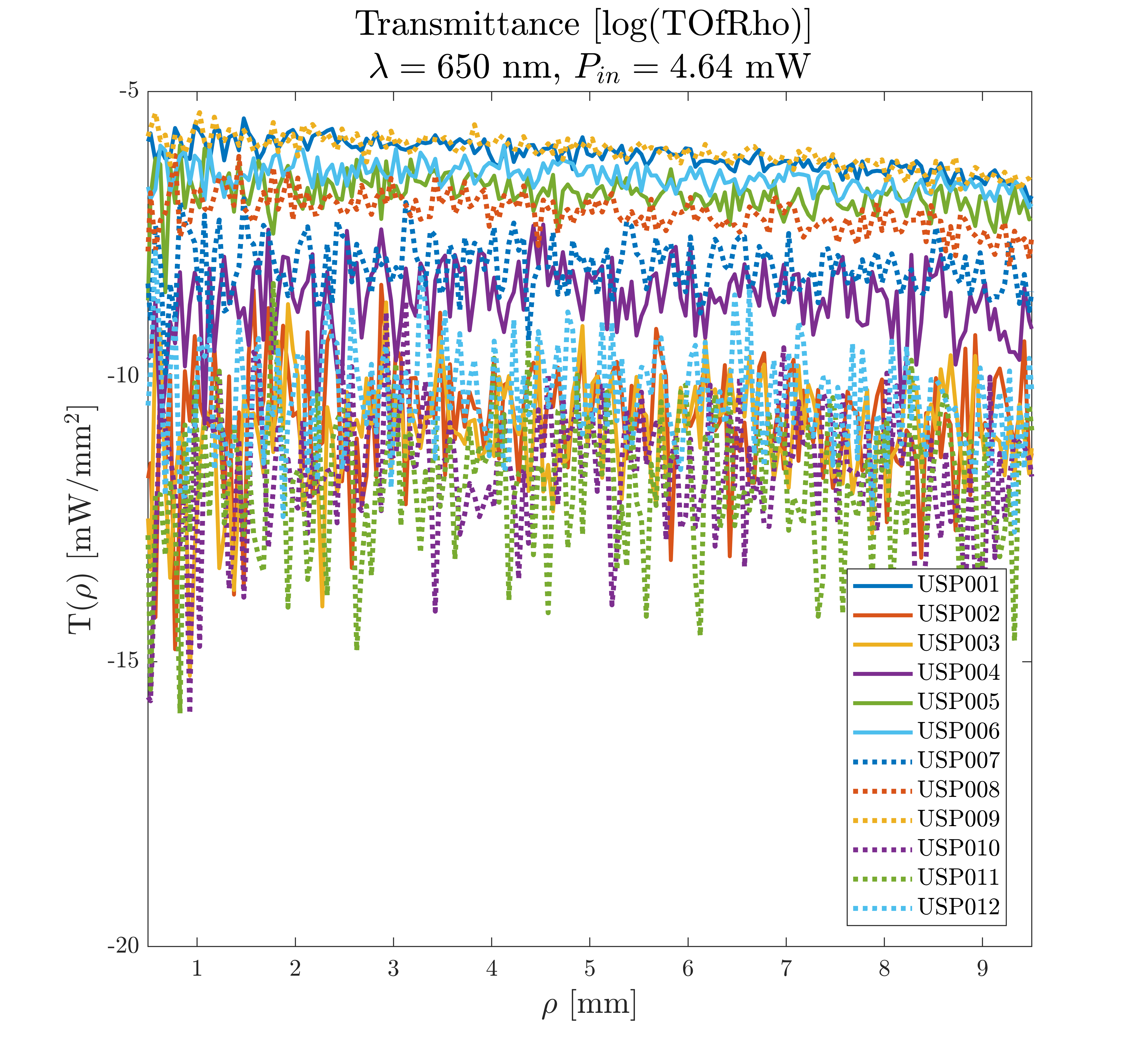}%
\includegraphics[width=0.49\columnwidth]{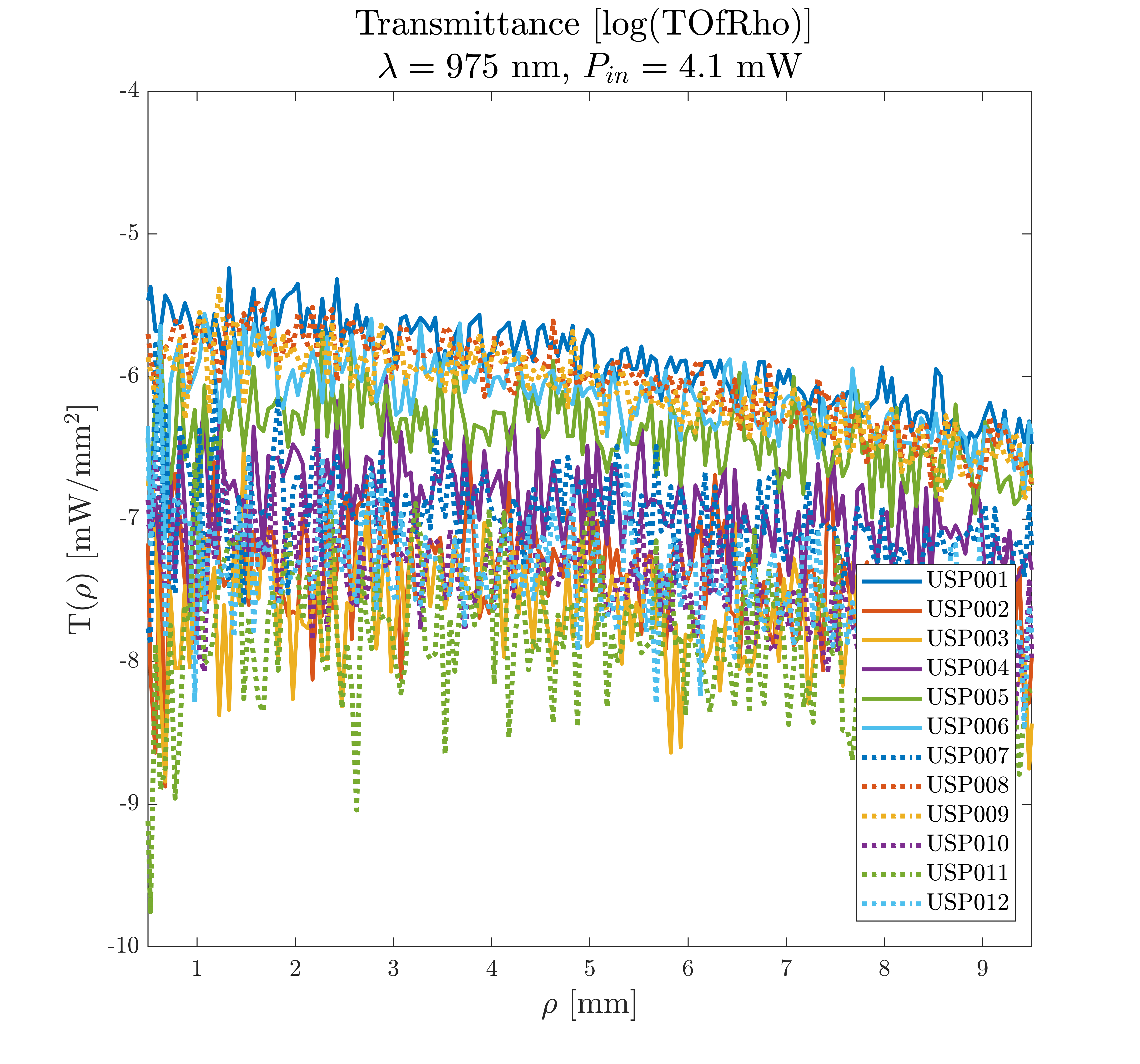}
\caption{The combination of all reflectance and transmission measurements using MCCL for all participants is presented. Clearly, there is a relationship between the skin type and the wavelengths when comparing these results. When considering the skin type and the wavelength, skin types I, II and near-infrared wavelength show the lowest degrees of absorption in the tissue bulk. Refer to \cite{Kallepalli_USMCData_2020} for complete set of results including simulations when using 750 nm and 830 nm wavelengths.}
\label{fig:MCCLResults_Transm_Ref_Results}
\end{figure*}

Considering the case of using photons at 650 nm wavelength, four distinct groupings emerge for reflectance data. USP005, USP006 and USP009 have a higher degree of reflectance; USP001, USP007 and USP008 in the second group; USP004 distinctly reflects lower optical energy and the remaining can be categorised with the lowest photons reflected. The greater transmission due to lower melanin concentration in the epidermis is clearly seen as USP001, USP005, USP006 and USP009 (Skin types I, II; Table \ref{tab:DepthThickUS}). Further, amongst these, USP001 anatomically has a thicker dermis layer which results in higher absorption. Participant USP011 has the thickest of optical models and the highest degree of melanin concentration in the epidermis layer resulting in the lowest degree of transmission. The thinner of models (USP001, USP009) with skin type I results in a higher degree of transmission. Photons at longer wavelengths travel further into biological tissue due to lesser attenuation. This general observation has been validated in previous literature \cite{Kallepalli2019_1,Kallepalli2020}. This is due to reduced attenuation due to melanin, resulting in a greater degree of penetration into the biological tissue. An individual assessment of reflectance and transmittance for each participant model at all wavelengths allows the understanding of energy transport through the tissues. Fluence and absorption results are shared in the CORD repository \cite{Kallepalli_USMCData_2020}. 

\begin{figure*}[t]
\centering
\includegraphics[width=0.49\columnwidth]{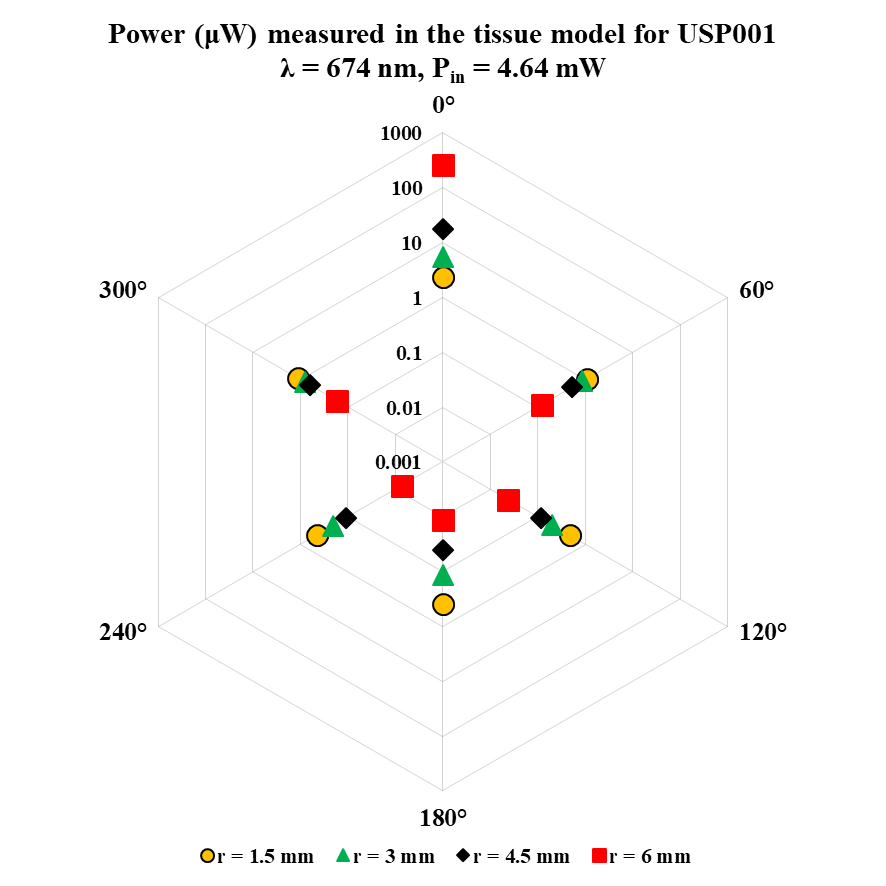}
\includegraphics[width=0.49\columnwidth]{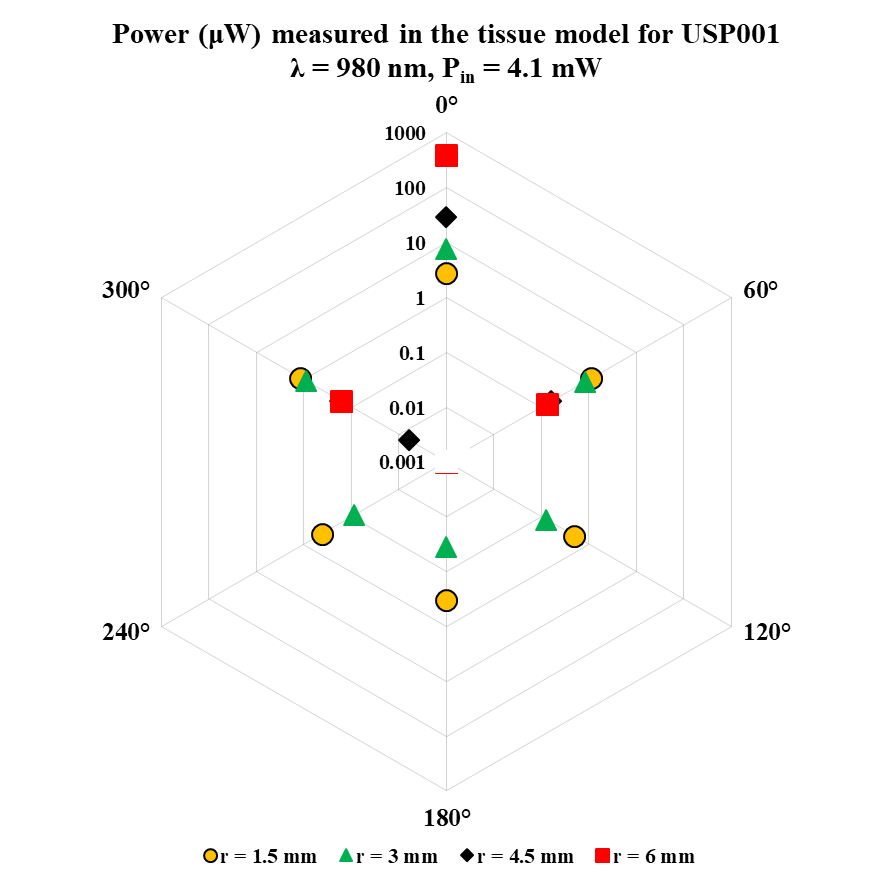}%
\\
\includegraphics[width=0.49\columnwidth]{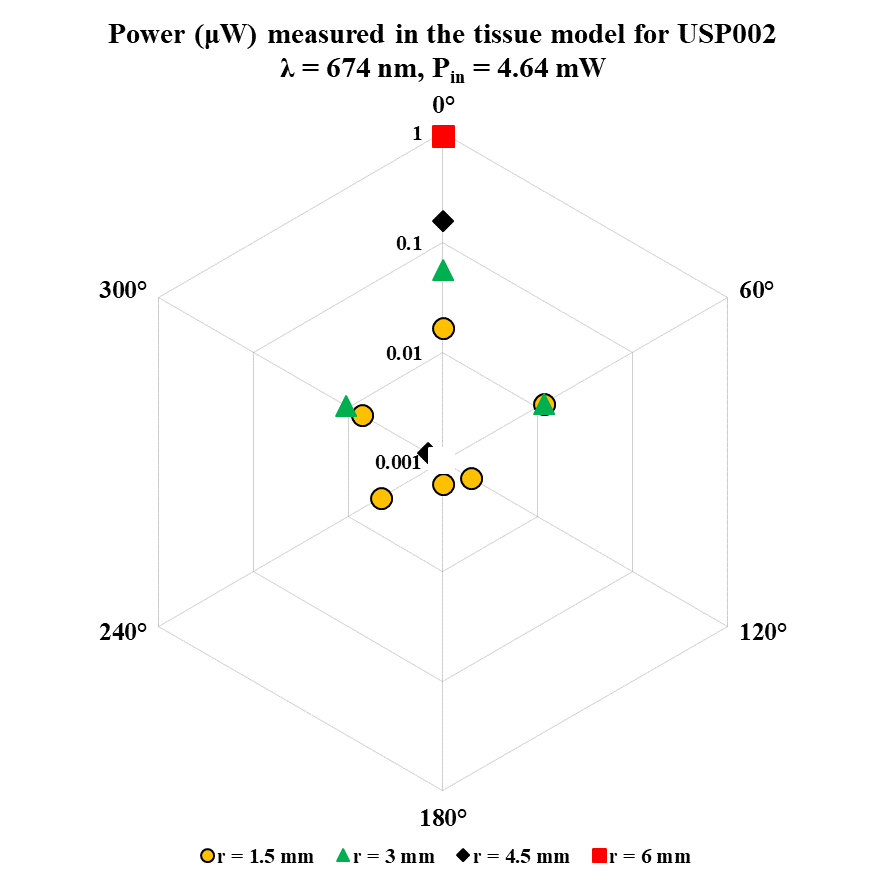}
\includegraphics[width=0.49\columnwidth]{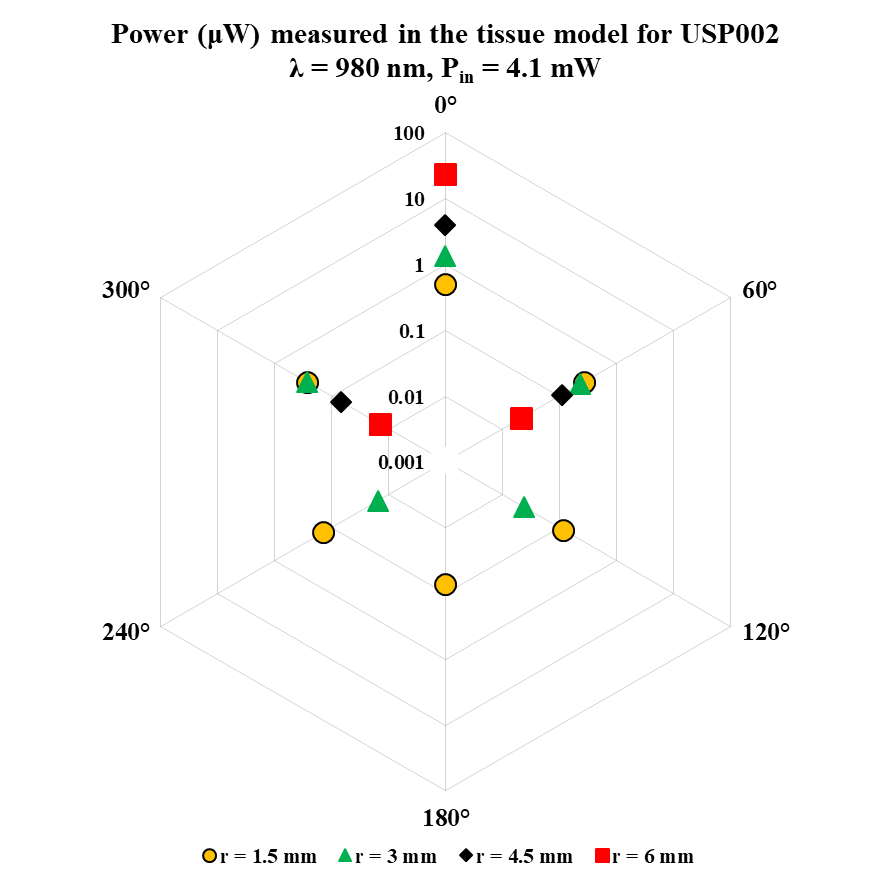}%
\caption{The radar plots illustrate the power detected at different depths for both wavelengths for participants USP001 and USP002. The plots illustrate power measured deeper in the tissue when using near-infrared wavelengths as opposed to red in this instance. The plots also illustrate the influence of different melanin concentrations in the epidermis layer, influenced by a greater degree of absorption. The orientation of the plots are analogous to Figure \ref{fig:TraceProUSMC_Model12} (\textbf{A}). Refer to \cite{Kallepalli_USMCData_2020} for complete set of results including simulations when using 750 nm and 830 nm wavelengths.}
\label{fig:TraceProResults_RadarPlots}
\end{figure*}

The reflection and transmission at the surfaces of the model are also quantified with dedicated detectors. The combination of all the participants' results at two wavelengths (red, near-infrared) as a function of increasing distance from the normal axis, i.e. the source position (Figure \ref{fig:MCCLResults_Transm_Ref_Results}). For instance, the photons travel further in the tissue when illuminated longer wavelengths although absorption is higher for skin type III, IV due to higher melanin distribution. Realistically, the amount of energy detectable with increasing source-detector separation reduces exponentially (within the scope of this study). The most important information, in this instance, is usually within 1-2 mm from the source. When comparing the participants' model at specific wavelengths, trends of absorption and overall reflectance emerge. These trends depend on a combination of thickness of the layers, optical properties of the tissue and the interaction of the photons at each instance. Alternatively, individual-specific measurements at all wavelengths were compared and can be found in the online repository \cite{Kallepalli_USMCData_2020}. 

To summarise the MCCL results, the highest reflectance was observed with photons at 975 nm wavelength. Reflectance typically involves photons that have interacted with the first two or three layers before being scattered back in the direction of the detector (which is in the same plane as the source). Measuring reflectance from any of the full anatomy models can be best achieved close to the sensor. In all the results, measurable and substantial reflectance is seen when $\rho \leq 1$ mm. The reflectance trend across all wavelengths changes drastically when the influence of melanin in the layers is considered. The difference in the amount of reflectance when comparing the wavelengths is directly proportional to the amount of melanin in the epidermis. For example, the difference between 650 nm and 975 nm reflectance is lesser for USP001 in comparison to USP012. The absorption of photons at 975 nm due to water in the tissue causes the rapid decrease in reflectance results at $\rho \geq 4$ mm for skin types IV--VI. For lower melanin volume fractions in skin types I and II, this decrease is less prominent as attenuation due to melanin is comparatively low. The transmittance of 675 nm photons through the tissue layers is the lowest amongst all wavelengths for all models. The absorption due to melanin at any volume fraction concentration is dominant, in addition to absorption due to haemoglobin, blood and water. Collagen fibres mainly scatter the light isotropically at the depths at which muscles are present, with water being the primary absorber in the bone and muscle tissues. This conclusion is supported by the research investigating the various factors of attenuation in the tissues considered in this study \cite{Jacques2013,Alexandrakis2005,Kienle1996,Ugnell1997,Ascenzi1959,KonugoluVenkataSekar2016}. 

Using MCCL, the results of this methodology showed that the reflectance and transmission through a tissue model and the transport of energy in the tissue volume can be analysed. The primary factors influencing this are the distribution of chromophores and the thickness of the layers. These are the most important requirements when developing a strategy for photodynamic therapy. Using ultrasound imaging data, we further wanted to explore the distribution of light within tissue volume with the effective construction of models true to the geometry of the finger. For this, TracePro was chosen; a proprietary package with an improved graphical user interface and the ability to add 3D geometries. Further, we can incorporate the blood vessels in the optical mode by using the depth measurements from the ultrasound data. Therefore, when using wavelengths that can penetrate the skin layers, an understanding of the distribution of energy within the tissue volume is valuable. To illustrate the results from TracePro, radar maps are constructed from the measurements of the radially distributed detectors (Figure \ref{fig:TraceProUSMC_Model12}\textbf{A}). The wavelengths used to replicate the laboratory lasers and emit collimated beams at 674 nm and 980 nm. The results at different depths and orientations from positions correspond to the detectors seen in the cross-section of the model. The absence of a marker in the plot signifies that no energy was detected at that position. 

\section{Conclusions} \label{sec:conclusions}
Using relatively inexpensive and non-invasive ultrasound imaging improves the specific to each individual, thereby improving the chances of successfully destroying the tumour cells. The novelty of this methodology is that it eliminates the assumptions of the tissue layer geometries and allows individual-specific modelling of possible treatment strategies. 

HFUS can conclusively resolve and provide adequate information regarding the definition of tissues and their dimensions for individual-specific analysis. The cost and advantage of having this information far outweighs any difficulty in integrating this step into current protocols. The methodology reduces the ambiguity associated with the trial and error nature of PDT and photomodulation treatments that are currently used for squamous cell carcinoma, basal cell carcinoma and other types of skin cancers. This method can provide conclusive evidence of the dimensions of the superficial tumour using a non-invasive and instantaneous method. This is important in itself as treating these cancers early improves the chances of survival than after they metastasise to other locations in the body.


\section*{Acknowledgments}
The authors would like to thank Dr Elizabeth Price, Mrs Catherine Lewis-Clarke and the fantastic staff of the Department of Radiology at the Great Western Hospital (Swindon, UK) for their assistance and support through the course of the research collaboration. The ultrasound scans and the value added to this article would not have been possible without their enthusiasm and support. We would also like to acknowledge Dr Elisa Barcaui (Clinica Dermatol\'ogica de Ipanema, Brazil), Dr Vincent Chan (University of Toronto), Dr Jeffrey Ketterling (Riverside Research, New York), Dr Ellen Marmur (Marmur Medical, New York), Dr Ronald H. Silverman (Columbia University, New York) for their correspondence and advice. Furthermore, we acknowledge the advice from Dr Brandon Corbett, Dr Constantinos Franceskides and Ms Am\'elie Grenier for designing, collaborating and debugging advice. 

This work was made possible through open-source software resources offered by the Virtual Photonics Technology Initiative, at the Beckman Laser Institute, University of California, Irvine.

\section*{Author contributions}
The research was conducted while AK was at Cranfield University. The research was conceptualised by AK, DBJ and MAR. The participant interaction and experimental procedures (data acquisition, processing and analysis) were carried out by AK and DBJ. The ultrasound scans and data analysis was done by JH. This research article was drafted by AK, with input and assistance from the remaining three authors. All the authors have read this article before submission for peer review. 

\section*{Financial disclosure}
None reported.

\section*{Conflict of interest}
The authors declare no potential conflict of interests.

\section*{Data Availability Statement}
The detailed results and accompanying information is available on the Cranfield Online Research Depository (CORD)\cite{Kallepalli_USMCData_2020}.

\bibliography{sample}%

\subsection*{Graphical Abstract Text}
Diagnostic medicine has benefited enormously from ultrasound imaging as a point-of-care device. In this research, we use high frequency ultrasound imaging to build optical models with individual-specific geometries and optical properties. Subsequently, these are used to model the interactions of light with Monte Carlo methods. The approach is intended to improve the effectiveness of light therapies with the flexibility testing a strategy prior to treatment. 

\subsection*{Graphical Abstract Figure}

\begin{figure}[htp]
\centering
\includegraphics[width=5cm]{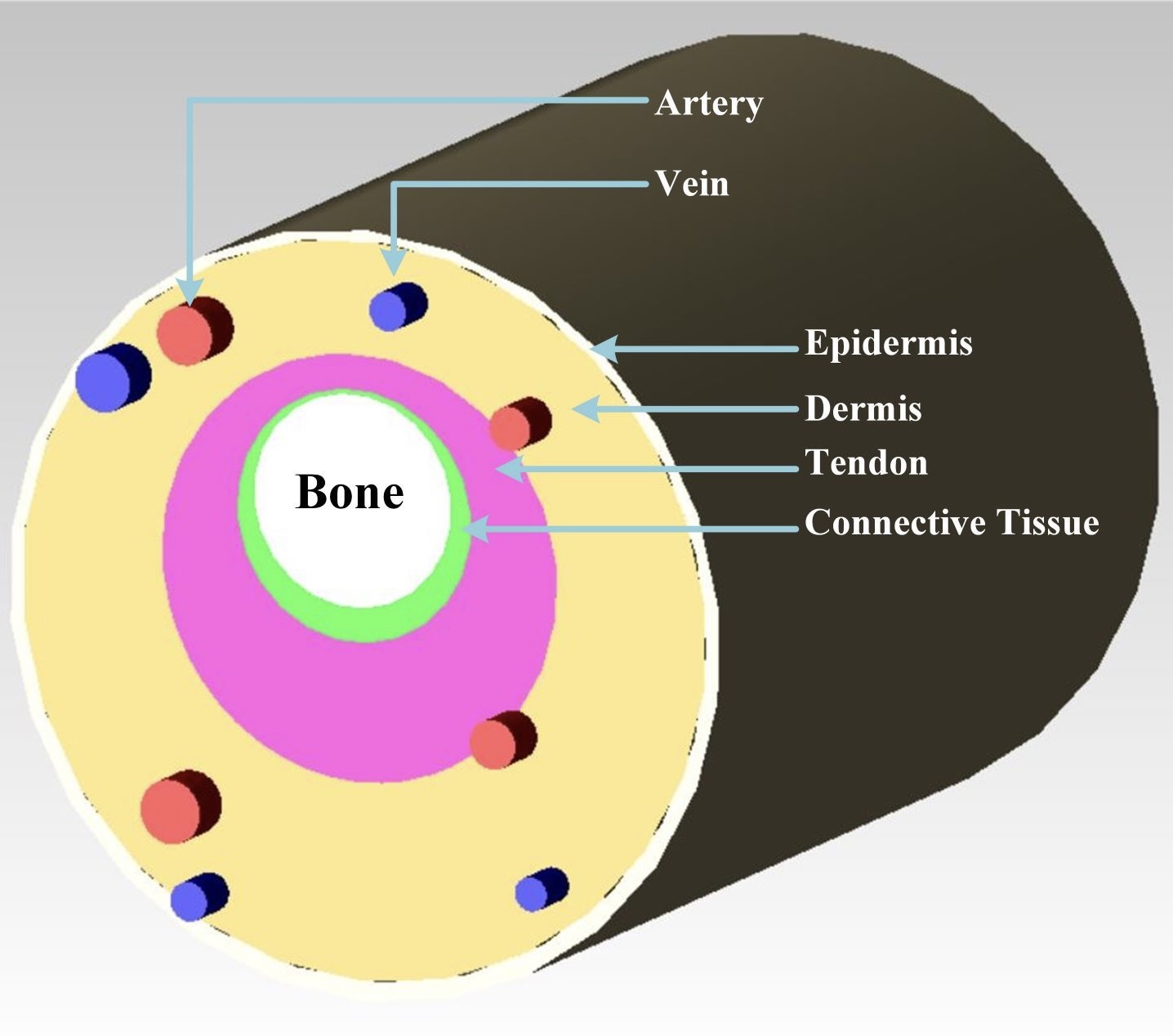}%
\end{figure}

\end{document}